\documentclass[prd,aps,showpacs,preprintnumbers,amssymb]{revtex4}

\usepackage{epsf}

\begin{document}

\title{%
\hfill{\normalsize\vbox{%
\hbox{\rm SU-4252-875}
}}\\
{
Global aspects of the scalar meson puzzle
}}

\author{Amir H. Fariborz $^{\it \bf a}$~\footnote[3]{Email:
fariboa@sunyit.edu}}

\author{Renata Jora $^{\it \bf b}$~\footnote[2]{Email:
cjora@physics.syr.edu}}

\author{Joseph Schechter $^{\it \bf c}$~\footnote[4]{Email:
schechte@physics.syr.edu}}

\affiliation{$^ {\bf \it a}$ Department of Mathematics/Science,
State University of New York Institute of Technology, Utica,
NY 13504-3050, USA.}

\affiliation{$^ {\bf \it b}$ INFN Roma, Piazzale A Moro 2,
Roma, I-00185 Italy.}

\affiliation{$^ {\bf \it c}$ Department of Physics,
Syracuse University, Syracuse, NY 13244-1130, USA,}

\date{\today}

\begin{abstract}
    A generalized linear sigma model for low energy QCD
is employed to study the quark structure
 of eight low lying scalar isomultiplets
as well as eight low lying pseudoscalar isomultiplets.
The model, building on earlier work, assumes the possible 
mixing of
 quark anti-quark states
with others made of two quarks and two antiquarks.  
 No {\it a priori} assumption is made about the
 quark contents of the states, which emerge as  
predictions. An amusing and contrasting
pattern for the quark structure is found;
the lighter conventional pseudoscalars are,
 as expected, primarily of
two quark type whereas the lighter scalars
 have very large four quark admixtures. The
 new feature of the present paper compared
to earlier ones in this series involves
 the somewhat subtle 
and complicated
effects of SU(3) flavor breaking. They do not alter
the general pattern of two quark vs. four quark
mixing obtained in the SU(3)symmetric case but, of
course, give a more detailed picture.
\end{abstract}


\maketitle

\section{Introduction}

    Although it is an ancient subject, the study of 
light scalar mesons has greatly intensified in the last
generation. Some representative earlier works of this
period are given in 
\cite{vanBev}-\cite{BFMNS01}.

   The ``global" aspect of the scalar puzzle is
 the unusual spectroscopy of the light scalar nonet.
 At present, the scalars below 1 GeV
appear to fit into a nonet as:
\begin{eqnarray}
I=0: m[f_0(600)]&\approx& 500\,\,{\rm MeV}
\nonumber \\
I=1/2:\hskip .7cm m[\kappa]&\approx& 800 \,\,{\rm MeV}
\nonumber \\
I=0: m[f_0(980)]&\approx& 980 \,\,{\rm MeV}
\nonumber \\
I=1: m[a_0(980)]&\approx& 980 \,\,{\rm MeV}
\label{scalarnonet}
\end{eqnarray}
This level ordering is seen to be flipped
 compared to that
 of the standard vector
meson nonet:
\begin{eqnarray}
I=1: m[\rho(776)]&\approx& 776\,\,{\rm MeV}\hskip .7cm
n{\bar n}
\nonumber \\
I=0: m[\omega(783)]&\approx& 783 \,\,{\rm MeV}\hskip .7cm
n{\bar n}
\nonumber \\
I=1/2: m[K^*(892)]&\approx& 892 \,\,{\rm MeV}\hskip .7cm
n{\bar s}
\nonumber \\
I=0: m[\phi(1020)]&\approx& 1020 \,\,{\rm MeV}\hskip .7cm
s{\bar s}
\label{vectornonet}
\end{eqnarray}
Here the standard quark content (n stands for a
 non-strange quark while s stands for a strange
quark) is displayed at the end for each case. The
vector mass ordering is seen to just correspond
 to the number of s-type quarks in each state. 
 It was pointed out a long time ago in Ref. \cite{j},
that the level order is automatically flipped
when mesons are made of two quarks and two antiquarks
instead of a single quark and antiquark.
That argument was given for a diquark- anti diquark
structure but is easily seen to also hold for a 
meson-meson ``molecule" type structure which was advocated,
 at least for a
partial nonet, in Ref. \cite{iw}. Note
that, in the ``ideal" four quark picture, the states in
 Eq.(\ref{scalarnonet}) consecutively have
 the quark contents: $nn{\bar n}{\bar n}$,
 $nn{\bar n}{\bar s}$, $ns{\bar n}{\bar s}$
and $ns{\bar n}{\bar s}$.
     One also notes that
the masses of the putative scalar nonet are
 significantly
lower than the other (tensor and two axial vector)
p-wave quark-antiquark nonets. There are enough other
scalar candidates [a$_0$(1450), K$_0$(1430) and 
two of 
f$_0$(1370), f$_0$(1500,) f$_0$(1710)] to make another
nonet although the masses of its
contents seem somewhat higher than an 
expected
scalar p-wave nonet. Based on the usual effect that
two mixing levels repel as well
as some more detailed features, it was suggested 
\cite{BFS3} that a global picture of these scalars
might consist of a lighter ``four quark" nonet mixing
with a heavier ``two quark" nonet. Further work
in this direction has been presented by a
 number of authors \cite{mixing}-\cite{thermo}. 

    A field theoretic toy model to study these features
was introduced in \cite{sectionV}. Although simple
 in conception
it turned out to be rather complicated
 to implement; as a result
the present authors examined its features in further detail
in \cite{FJS05,1FJS07,2FJS07,3FJS07,FJS08}. While the general 
treatment in \cite{FJS05} only used the tree level Ward-type 
identities of the theory, the treatments in \cite{1FJS07}
and \cite{2FJS07} employed specific potential terms chosen in 
a systematic way to correspond to terms with the
minimum number of underlying quark lines. 
 Furthermore the quark mass terms were 
neglected, which not only is a simplifying feature but is also
a check that the peculiar mixing results found are not an
artifice of a particular choice of symmetry breaking terms. 
(In QCD it is expected that the main features of the various 
particle multiplets should also hold in the limit of zero 
light quark masses.) Next, in \cite{3FJS07}
it was demonstrated that the results did not change much when
the minimal mass term was included with equal
masses for all three light quarks. It was also shown that
the current algebra theorem \cite{w}
 for pion pion scattering 
holds to
a very good approximation in this more complicated case where
the ordinary pion has a small admixture of a state containing 
two quarks and two antiquarks. The theorem has a very small
correction due to a very small violation of the partially
conserved axial vector current hypothesis in the model, 
however. Most recently, in \cite{FJS08} an amusing connection
between the model and the instanton approach to QCD dynamics
was discussed. This connection is not so surprising when
one notices that 
it is the (broken by instantons) U(1)$_A$ symmetry which 
formally distinguishes the ``four quark" from
 the ``two quark" meson states.

   In the present paper we include the flavor
SU(3) symmetry breaking in the model. As will
be seen, this is not a very simple matter, even
in the tree approximation being used. Within this
framework we will not
introduce any further
approximations in our numerical treatment.
Furthermore, the sensitivity of the predictions to
the main uncertainties in the experimental
inputs will be explicitly displayed.

   Considering the relatively large number of 
``outputs," it may be desirable to immediately
 just display
our main ``typical" results. These are the 
masses and the ``two quark" vs. 
``four quark" percentages of the members of
all four nonets (light and heavy pseudoscalars
and light and heavy scalars). They are listed
in Tables \ref{phi_content_1215}  and  
\ref{s_content_1215}. Isospin but not SU(3)
symmetry is being assumed. Note that for the 
I=1/2 and I=1 states, the prime denotes the heavier
particle. For the I=0 particles there are four
states of each parity and they are denoted 
by subscripts 1, 2, 3, 4 in order of increasing 
mass.
Altogether, considering the isospin degeneracy,
there are 16 different masses. The 8 inputs
comprise the pion decay constant, the four masses:
[$m_\pi$, $m_{\pi'}$, $m_a$, $m_a'$ ], the 
strange to
non strange quark mass ratio (which is related to
assuming a value for $m_K$) and (as to be 
explained later) the sum and the product of all
the four I=0 pseudoscalar squared masses for each of 
the possible scenarios for their identification
with experimental states.

    Thus there are 9 mass predictions and 16
predictions for the two-quark and four quark 
percentages being displayed in 
Tables \ref{phi_content_1215}  and
\ref{s_content_1215}. To compare with 
Eq.(\ref{scalarnonet}) for the
``experimental" light scalar nonet,
we read off the mass pattern:
\begin{equation}
f_1(742),\, \kappa(1067), \, a(980)
\, f_2(1085),
\label{lightscalarpred}
\end{equation}
in which only the $a(980)$ mass was
an input. Clearly, the pattern,
featuring an approximate degeneracy of one 
I=0 state and the I=1 state
together with a flipping compared to
 the standard
vector meson order, is very
 similar to  the experimental one
in Eq.(\ref{scalarnonet}).
Actually, these ``tree level" predictions
are expected to receive some
non-trivial corrections as to be discussed
later. We also read from 
Table \ref{s_content_1215} that the light 
scalar nonet is predicted to be
 predominantly of 4-quark type. On the other hand,
the heavy scalar nonet is predicted to be 
predominantly of 2-quark type and to have the
standard (vector meson like) pattern:
\begin{equation}
a'(1474),\,f_3(1493),\,\kappa'(1624),\,
f_4(1784).
\label{heavyscalarpred}
\end{equation}

   The conventional light 
pseudoscalar nonet is read from Table 
\ref{phi_content_1215} as,
\begin{equation}
\pi(137), \,
K(515), \, \eta_1(553),\,\eta_2(982),
\label{lightpsnonet}
\end{equation}
where the structure is clearly
 close to the conventional one.
Notice the well known fact that this
nonet agrees in ordering with that of
 the standard
vector nonet in the sense that the I=1
state is lightest but disagrees in that
the lighter I=0 state is not at all
close to the I=1 state.
 At the theoretical level,
this arises from axial U(1) (instanton 
type) terms as are included here. It can also 
be read off that the light pseudoscalar
nonet members are predicted to be predominantly 
 of ``2-quark", i.e.
${\bar q}q$, type. In contrast, the 
heavy pseudoscalar nonet members are
predicted to be of predominantly 
``4-quark" type with a similar
  mass ordering 
as the light pseudoscalars; namely,
\begin{equation}
\pi'(1215), \,
K'(1195), \, \eta_3(1225),\,\eta_4(1784).
\label{heavypsnonet}
\end{equation}
In this multiplet, the distortion of the
pattern from that of the standard vector
multiplet seems to reflect the greater role
of ``instanton" effects over
``4-quark" effects. 

   A detailed discussion of how these
 results were obtained is given in the following
sections. Section II briefly summarizes the ``toy
Lagrangian" and the choice of input parameters.
The needed mass matrices are presented
 in section III and Appendix B. The first
part of Section IV and, especially, Appendix A
explain in a step by step way how the 
Lagrangian parameters (other than those
associated with the ``instanton" terms) are
related to the experimental data. The 
second part of section IV explains predictions of
the model and their sensitivity to changes
in the input parameters. Section V explains the
work associated with the instanton terms
and the complicated I=0 pseudoscalar sector.
Finally, conclusions and some further discussion
are presented in section VI.

\begin{table}[htbp]
\begin{center}
\begin{tabular}{c|c|c|c}
\hline \hline
State & \,\, ${\bar q} q$\% \, \, & \, ${\bar q} 
{\bar q} q q$\%\, & \, $m$ (GeV)
\\
\hline
\hline
$\pi$      & 85  & 15 & 0.137\\
\hline
$\pi'$      & 15  & 85 & 1.215\\
\hline
$K$          & 86  & 14 & 0.515\\
\hline
$K'$          & 14  & 86 & 1.195\\
\hline
$\eta_1$   &  89   &  11   & 0.553 \\
\hline
$\eta_2$   &  78   &  22   & 0.982\\
\hline
$\eta_3$   &  32   &  68   &  1.225 \\
\hline
$\eta_4$   &  1   & 99     & 1.794\\
\hline
\hline
\end{tabular}
\end{center}
\caption[]{
Typical predicted properties of pseudoscalar
states:
${\bar q} q$ percentage (2nd
column), ${\bar q} {\bar q} q q$ (3rd
column) and masses (last column).}
\label{phi_content_1215}
\end{table}

\begin{table}[htbp]
\begin{center}
\begin{tabular}{c|c|c|c}
\hline \hline
State &\,\, ${\bar q} q$\%\,\,& \, ${\bar q} 
{\bar q} q q$\% \, & \, $m$ (GeV)
\\
\hline
\hline
$a$        &  24 &  76 & 0.984   \\
\hline
$a'$        &  76 &  24 & 1.474   \\
\hline
$\kappa$  &   8  &  92  & 1.067  \\
\hline
$\kappa'$  &   92  &  8  & 1.624 \\
\hline
$f_1$     &  40    &  60  & 0.742  \\
\hline
$f_2$     &  5   &    95  & 1.085 \\
\hline
$f_3$     &  63   &   37  & 1.493 \\
\hline
$f_4$     &  93   &   7   & 1.783 \\
\hline
\hline
\end{tabular}
\end{center}
\caption[]{ Typical predicted properties of scalar 
states: 
${\bar q} q$ percentage (2nd 
column), ${\bar q} {\bar q} q q$ (3rd 
column) and masses (last column).
 }
\label{s_content_1215}
\end{table}

\section{Model Lagrangian and physical inputs}

   The model employs the 3$\times$3 matrix
chiral nonet fields:
\begin{equation}
M = S +i\phi, \hskip 2cm
M^\prime = S^\prime +i\phi^\prime.
\label{sandphi}
\end{equation}
The matrices $M$ and $M'$ transform in the same way under
chiral SU(3) $\times$ SU(3) transformations but 
may be distinguished by their different U(1)$_A$ 
transformation properties. $M$ describes the ``bare"
 quark antiquark scalar and pseudoscalar nonet fields while
$M'$ describes ``bare" scalar and pseudoscalar fields 
containing two quarks and two antiquarks. At the
symmetry level with which we are working, 
it is unnecessary to 
further specify the four quark field configuration.
The four quark field may, most generally,
 be imagined as some linear
combination of a diquark-antidiquark  and a 
``molecule" made of two quark-antiquark ``atoms".
 
The general Lagrangian density which defines our model is
\begin{equation}
{\cal L} = - \frac{1}{2} {\rm Tr}
\left( \partial_\mu M \partial_\mu M^\dagger
\right) - \frac{1}{2} {\rm Tr}
\left( \partial_\mu M^\prime \partial_\mu M^{\prime \dagger} \right)
- V_0 \left( M, M^\prime \right) - V_{SB},
\label{mixingLsMLag}
\end{equation}
where $V_0(M,M^\prime) $ stands for a function made
from SU(3)$_{\rm L} \times$ SU(3)$_{\rm R}$
(but not necessarily U(1)$_{\rm A}$) invariants
formed out of
$M$ and $M^\prime$.

 As we previously discussed \cite{1FJS07}, the
 leading choice of terms 
corresponding
to eight or fewer quark plus antiquark lines
 at each effective vertex
reads:
\begin{eqnarray}
V_0 =&-&c_2 \, {\rm Tr} (MM^{\dagger}) +
c_4^a \, {\rm Tr} (MM^{\dagger}MM^{\dagger})
\nonumber \\
&+& d_2 \,
{\rm Tr} (M^{\prime}M^{\prime\dagger})
     + e_3^a(\epsilon_{abc}\epsilon^{def}M^a_dM^b_eM'^c_f + h.c.)
\nonumber \\
     &+&  c_3\left[ \gamma_1 {\rm ln} (\frac{{\rm det} M}{{\rm det}
M^{\dagger}})
+(1-\gamma_1){\rm ln}\frac{{\rm Tr}(MM'^\dagger)}{{\rm 
Tr}(M'M^\dagger)}\right]^2.
\label{SpecLag}
\end{eqnarray}
     All the terms except the last two have been chosen to also
possess the  U(1)$_{\rm A}$
invariance.
The symmetry breaking term which models the QCD mass term
takes the form:
\begin{equation}
V_{SB} = - 2\, {\rm Tr} (A\, S)
\label{vsb}
\end{equation}
where $A=diag(A_1,A_2,A_3)$ are proportional to
  the three light quark 
masses.  
The model allows for two-quark condensates,
$\alpha_a=\langle S_a^a \rangle$ as well as
four-quark condensates 
$\beta_a=\langle {S'}_a^a \rangle$.
Here we assume \cite{SU} isotopic spin
symmetry so A$_1$ =A$_2$ and:
\begin{equation}
\alpha_1 = \alpha_2  \ne \alpha_3, \hskip 2cm
\beta_1 = \beta_2  \ne \beta_3
\label{ispinvac}
\end{equation}

 We also need the ``minimum" conditions,
\begin{equation}
\left< \frac{\partial V_0}{\partial S}\right> + \left< \frac{\partial
V_{SB}}{\partial
S}\right>=0,
\quad \quad \left< \frac{\partial V_0}{\partial S'}\right>
=0. 
\label{mincond}
\end{equation}

There are twelve parameters describing the Lagrangian and the
vacuum. These include the six coupling constants
 given in Eq.(\ref{SpecLag}), the two quark mass parameters,
($A_1=A_2,A_3$) and the four vacuum parameters ($\alpha_1
=\alpha_2,\alpha_3,\beta_1=\beta_2,\beta_3$).The four minimum
equations reduce the number of needed input parameters to
eight.
 
Five of these eight are supplied by the following 
masses together with the pion decay constant:
\begin{eqnarray}
 m[a_0(980)] &=& 984.7 \pm 1.2\, {\rm MeV}
\nonumber 
\\ m[a_0(1450)] &=& 1474 \pm 19\, {\rm MeV}
\nonumber \\
 m[\pi(1300)] &=& 1300 \pm 100\, {\rm MeV}
\nonumber \\
 m_\pi &=& 137 \, {\rm MeV}
\nonumber \\
F_\pi &=& 131 \, {\rm MeV}
\label{inputs1}
\end{eqnarray}
Because $m[\pi(1300)]$ has such a large uncertainty,
we will, as previously, examine predictions   
depending on the choice of this mass
within its experimental range.
The sixth input will be taken as the light 
``quark mass ratio" $A_3/A_1$, which will
be varied over an appropriate range.
 The remaining two inputs will be taken from the
 masses of the four (mixing) isoscalar, pseudoscalar 
mesons. This mixing is characterized by a 4 
$\times$ 4 matrix
$M_\eta^2$. A practically convenient choice is to consider
Tr$M_\eta^2$ and det$M_\eta^2$ as the inputs.

Given these inputs there are a very large number of 
predictions. At the level of the quadratic terms in the 
Lagrangian, we predict all the remaining masses
 and decay constants as well
as the angles describing the mixing between each of
($\pi,\pi'$),
($K,K'$), ($a_0,a_0'$), ($\kappa,\kappa'$) multiplets
and each of the 4$\times$4
isosinglet mixing matrices
 (each formally described by six angles).

Defining the total potential $V=V_0+V_{SB}$, the
four minimum conditions
explicitly read:

\begin{equation}
\left\langle 
{{\partial \, V} \over {\partial \, S_1^1} }
\right\rangle
 = 
4\,  e_3^a \,  \alpha_1  \,  \beta_3 + 
4 \,  e_3^a \, \beta_1 \, \alpha_3 - 2 \,  c_2 \,  \alpha_1 
+ 4\, c_4^a  \, \alpha_1^3 - 2 \,
A_1 = 0.
\end{equation}

\begin{equation}
\left\langle 
{{\partial \, V} \over {\partial \, S_3^3} }
\right\rangle
=   
8 \,  e_3^a \,  \beta_1 \,  \alpha_1 - 
2 \,  c_2 \,  \alpha_3 + 
4 \,  c_4^a \,  \alpha_3^3 - 2 \,  A_3 = 0.
\end{equation}

\begin{equation}
\left\langle 
{{\partial \, V} \over {\partial \, {S'}_1^1} }
\right\rangle
 = 
4 \,  e_3^a \, \alpha_1 \,  \alpha_3 + 2 \,  d_2 \,  
\beta_1 = 0.
\end{equation}

\begin{equation}
\left\langle 
{{\partial \, V} \over {\partial \, {S'}_3^3} }
\right\rangle
 = 
4 \,  e_3^a \,  \alpha_1^2 + 
2 \,  d_2 \,  \beta_3 = 0.
\end{equation}

\section{Mass matrices}

   In order to evaluate the eight
independent parameters from experiment it is necessary
to first obtain the formulas for the
 squared mass matrices corresponding to
 each set of particles with the same parity and 
isotopic spin quantum numbers. These are 
calculated by taking the second derivatives of the
potential with respect to the appropriate fields.

   For the two mixing ``pion" states (I=1, 
P=$-$) we find:

\begin{equation}
\left(  M_\pi^2  \right) =
 \left[ 
  \begin {array}{cc} 
4  \,  e_3^a  \,   \beta_3  -  2  \,  c_2  +  4  \,  
c_4^a  \,   \alpha_1^2
&
4  \,  e_3^a  \,  \alpha_3
\\
%-------------------------------------------------
4  \,  e_3^a  \,  \alpha_3
&
2  \,  d_2
\end{array}
 \right]
\end{equation}
 
The 2 $\times$ 2 K meson matrix (I=1/2, P=$-$) 
reads:

\begin{equation}
(M_K^2) =
 \left[ 
  \begin {array}{cc} 
4 \, e_3^a  \,  \beta_1  -  2  \,  c_2
+  4  \,  c_4^a  \,  \alpha_1^2  -  4  \,   c_4^a  \,
\alpha_1  \, \alpha_3  +  4
\,  c_4^a  \,  \alpha_3^2
&
4  \,  e_3^a  \,  \alpha_1
\\
%-------------------------------------------------
4  \,  e_3^a  \,  \alpha_1
&
2  \,   d_2
\end{array}
 \right]
\end{equation}

In the case of the two scalar a-mesons (I=1, P=+)
the mass squared mixing matrix reads: 

\begin{equation}
\left(X_a^2\right) =
\left[ 
\begin{array}{cc} 
-4  \,   e_3^a  \,  \beta_3 - 
2  \,   c_2  +  12  \,  c_4 \, 
\alpha_1^2
&
- 4  \,   e_3^a  \,  \alpha_3  
\\
-4  \,    e_3  \,   \alpha_3
&
2   \,   d_2
\end{array}
 \right]
\end{equation}

The final 2 $\times$ 2 mass squared matrix 
describes the
kappa-type scalars (I=1/2, P=+):

\begin{equation}
\left(  X_\kappa^2  \right) =
 \left[ 
  \begin {array}{cc} 
-4  \,   e_3^a  \,   \beta_1  -  
2  \,   c_2  +  4  \,   c_4  \,  \alpha_1^2
+ 4  \,   c_4^a  \,   \alpha_1  \,   \alpha_3  + 
4  \,   c_4  \,   \alpha_3^2
&
-  4  \,  e_3^a  \,  \alpha_1
\\
-  4  \,   e_3^a  \,  \alpha_1
&
2  \,   d_2
\end{array}
 \right]
\end{equation}

In the case of the I=0 scalars there are four
 particles which mix with each other; the squared
mass matrix then takes the form: 

\begin{equation}
\left(  X_0^2  \right) =
 \left[ 
  \begin{array}{cccc} 
4  \,   e_3^a  \,  {\it \beta_3}  -  2  \,  
c_2  +  12  \,   c_4^a  \,   \alpha_1^2
&
4  \, \sqrt{2}\,  e_3^a  \,  \beta_1
&
4  \,  e_3^a  \,   \alpha_3
&
4 \, \sqrt{2}\,    e_3^a  \,  \alpha_1
\\
%-------------------------------------------------
4  \, \sqrt{2}\,  e_3^a   \,  
\beta_1
&
-2  \,  c_2  + 12  \,  c_4^a  \, \alpha_3^2
&
4  \,  \sqrt{2}\, e_3^a  \,   
\alpha_1
&
0
\\
%-------------------------------------------------
4  \,  e_3^a  \,   \alpha_3
&
4  \,  \sqrt{2}\,  e_3^a   \, 
\alpha_1
&
2  \,   d_2
&
0
\\ 
%-------------------------------------------------
4  \,  \sqrt{2}\, e_3^a  \, 
\alpha_1
&
0
&
0
&
2 \, d_2
\end {array} 
\right]
\end{equation}
  
For this matrix the basis states are consecutively,

\begin{eqnarray}
f_a&=&\frac{S^1_1+S^2_2}{\sqrt{2}} \hskip .7cm 
n{\bar n},
\nonumber  \\ 
f_b&=&S^3_3 \hskip .7cm s{\bar s},
\nonumber    \\
f_c&=&  \frac{S'^1_1+S'^2_2}{\sqrt{2}}
\hskip .7cm ns{\bar n}{\bar s},
\nonumber   \\
f_d&=& S'^3_3
\hskip .7cm nn{\bar n}{\bar n}.
\label{fourbasis}
\end{eqnarray}
    The non-strange (n) and strange (s) quark content
for each basis state has been listed at the end of
each line above.

    In the case of the four I=0 pseudoscalars, the mixing
matrix formula is a long one which is given in Appendix B.

\section {Parameters and some predictions}

     A simplifying feature in the present model is that,
as previously,
the parameters other than $c_3$ and $\gamma_1$ may be
found without considering the four I=0 pseudoscalars.
This is due to the presence of ``ln's" in the last two terms
of the potential, $V_0$. So we consider the initial
 (six) parameters
first. In Appendix A, a method is presented for their 
consecutive determination in a convenient way.

\begin{figure}[t]
\begin{center}
\vskip 1cm
\epsfxsize = 6cm
\ \epsfbox{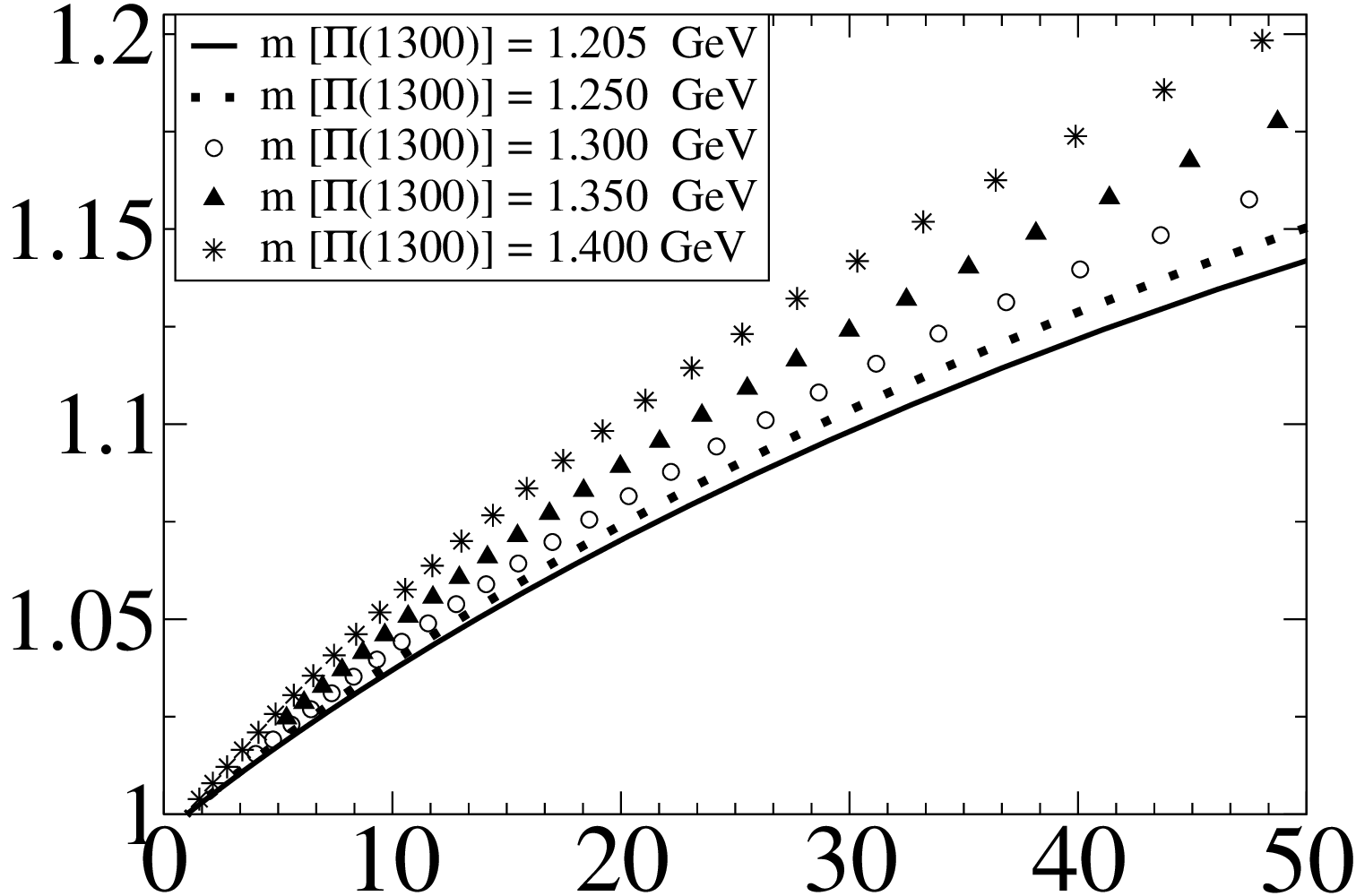}
\end{center}
\caption[]{%
$F_K\over F_\pi$ vs $A_3\over A_1$.
}
\label{Fig_Fkpi_vs_A31}
\end{figure}

 Here we 
consider the strange to non-strange quark mass ratio,
$A_3/A_1$ to be an input. For many years this has been known,
from various analyses of the chiral treatment
\cite{RPP}, of the light
pseudoscalars, to be around 25. In Fig.\ref{Fig_Fkpi_vs_A31}
the dependence of the predicted decay constant
 ratio $F_K/F_\pi$ on
the quark mass ratio, including the dependence on the 
parameter $m[\pi(1300)]$ is shown. For comparison, the
experimental ratio \cite{fratioexpt} is 
 $F_K/F_\pi\approx 1.19$. Thus, the model predictions
are seen to be quite reasonable, although perhaps a bit
too small for typical parameter choices. This ratio 
may be fine-tuned by including non-minimal kinetic
terms in the Lagrangian. See e.g. Eq.(4.1)
 of \cite{GJJS}.

\begin{figure}
\begin{center}
\vskip 1cm
%-----------------------------------------
\epsfxsize = 6cm
 \epsfbox{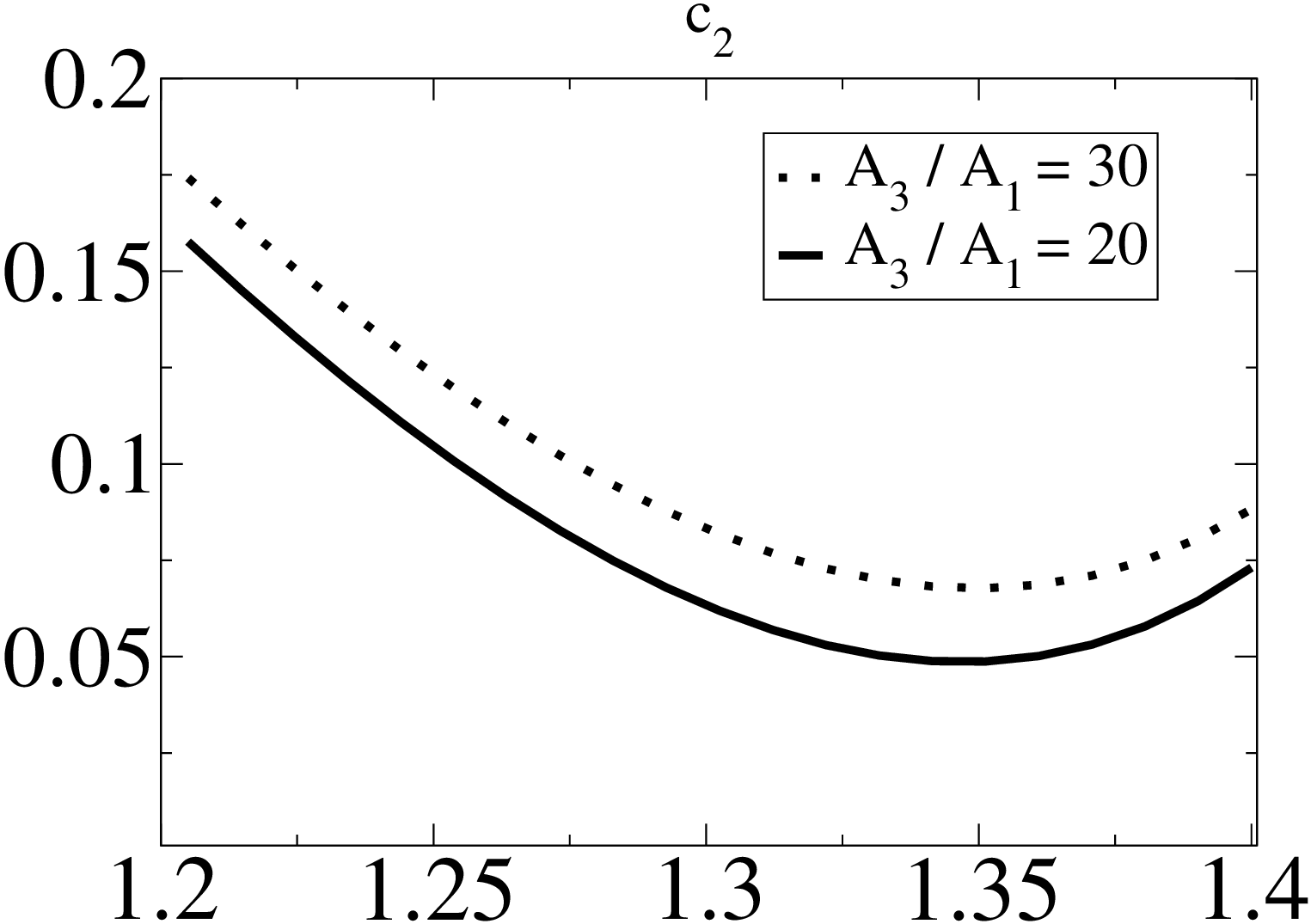}
\hskip 1cm
\epsfxsize = 6cm
 \epsfbox{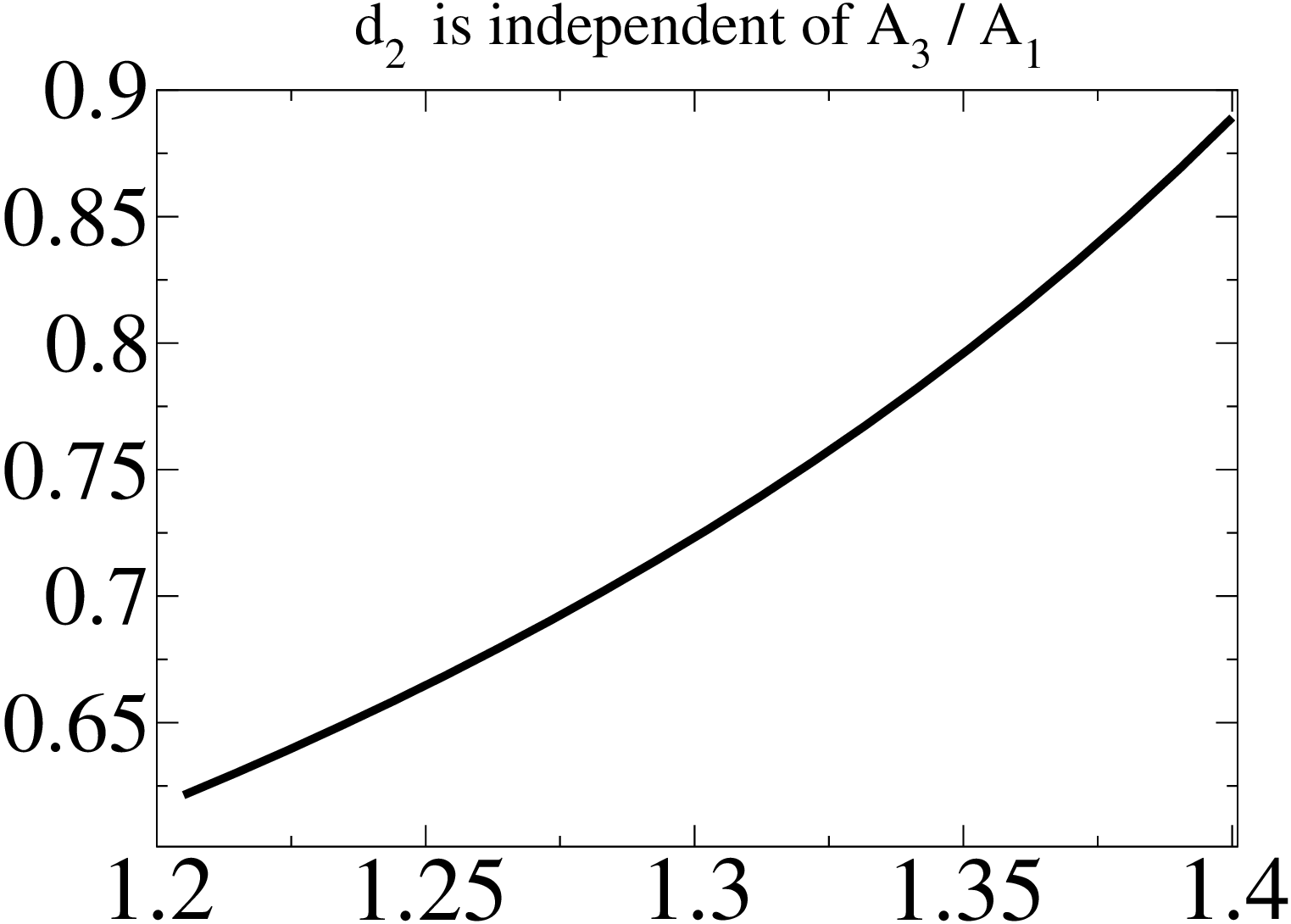}
\vskip 1cm
%-----------------------------------------
\epsfxsize = 6cm
 \epsfbox{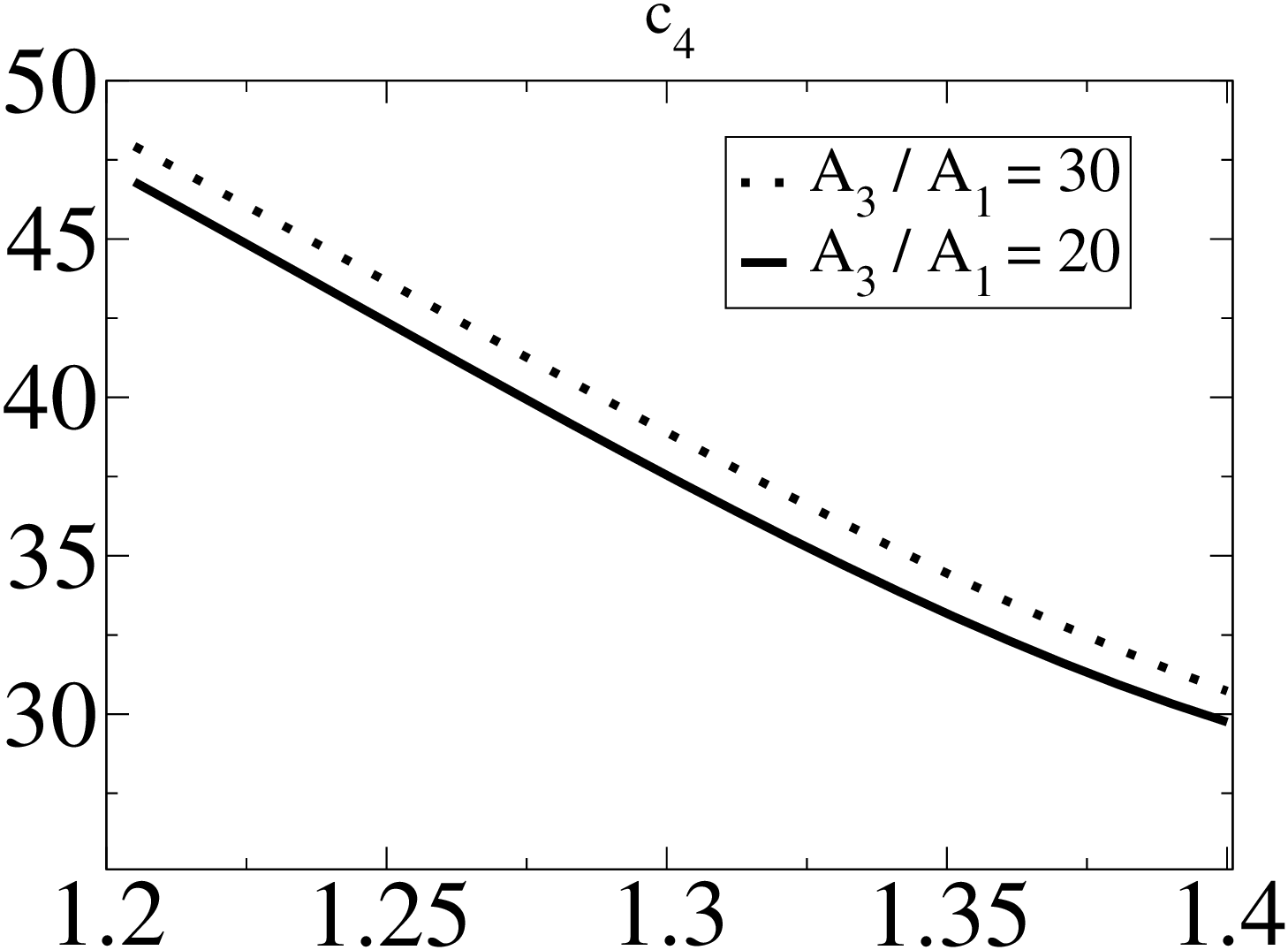}
\hskip 1cm
\epsfxsize = 6cm
 \epsfbox{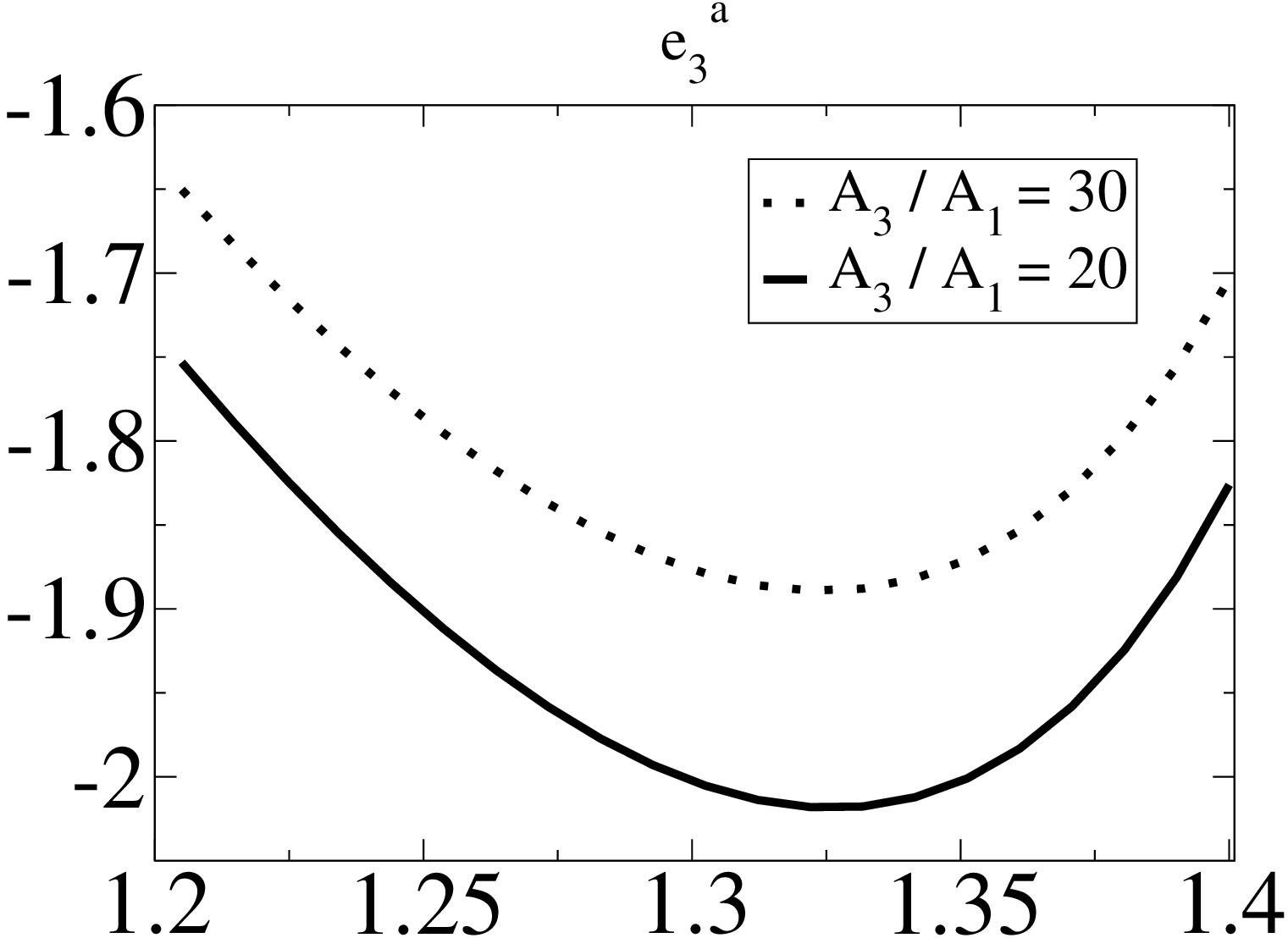}
\vskip 1cm
%-----------------------------------------
\epsfxsize = 6cm
 \epsfbox{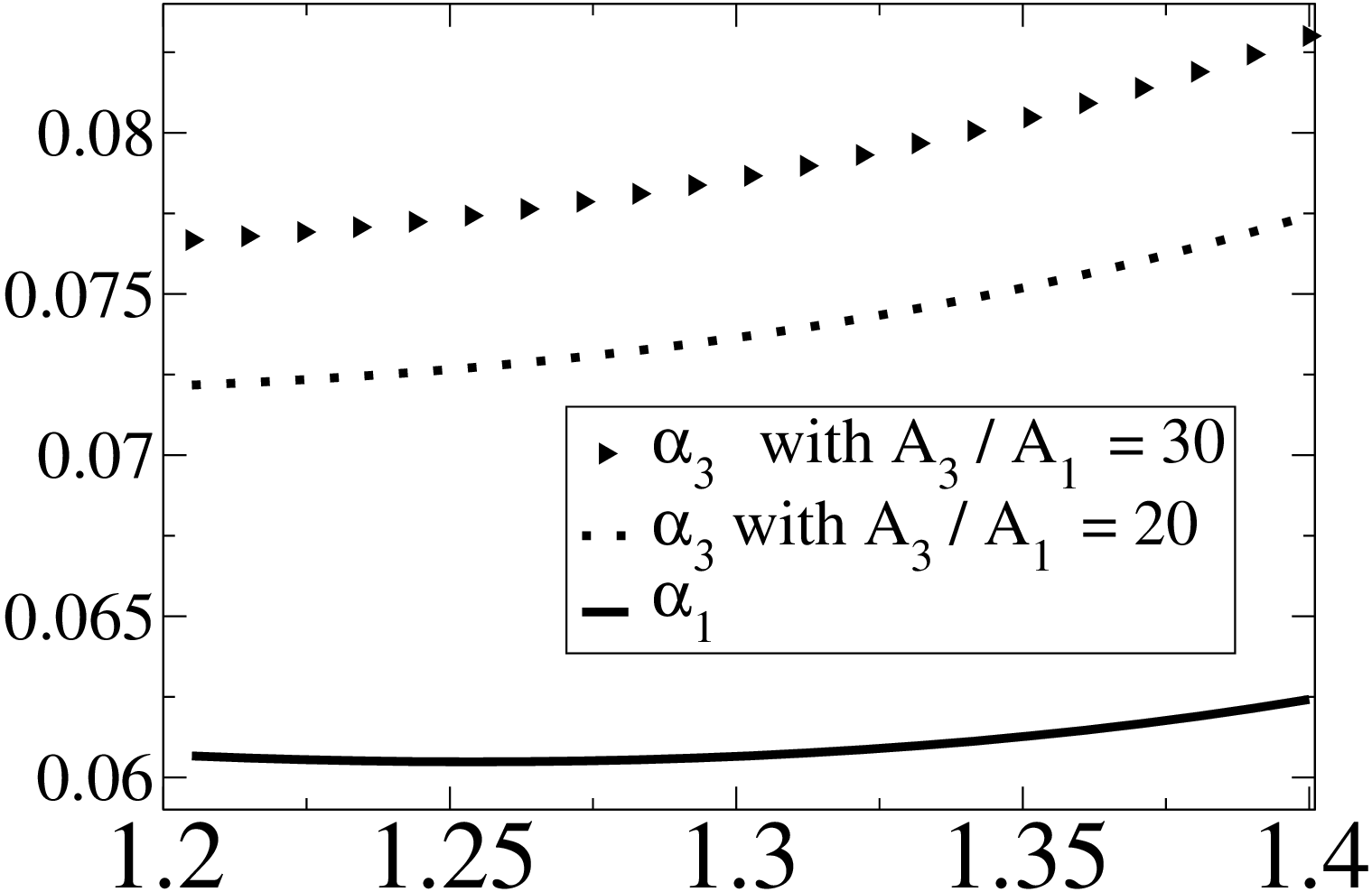}
\hskip 1cm
\epsfxsize = 6cm
 \epsfbox{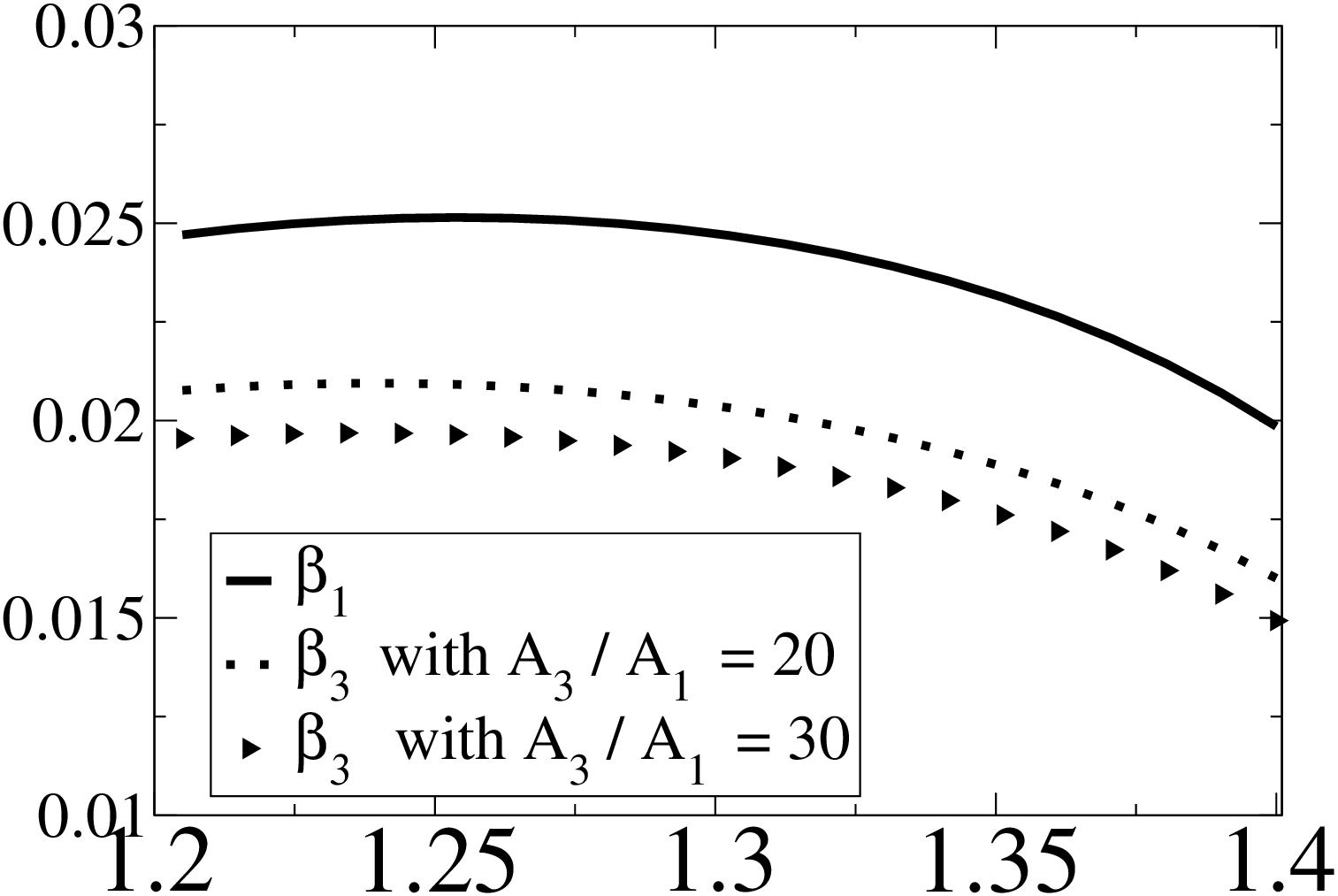}
\vskip 1cm
%-----------------------------------------
\epsfxsize = 6cm
 \epsfbox{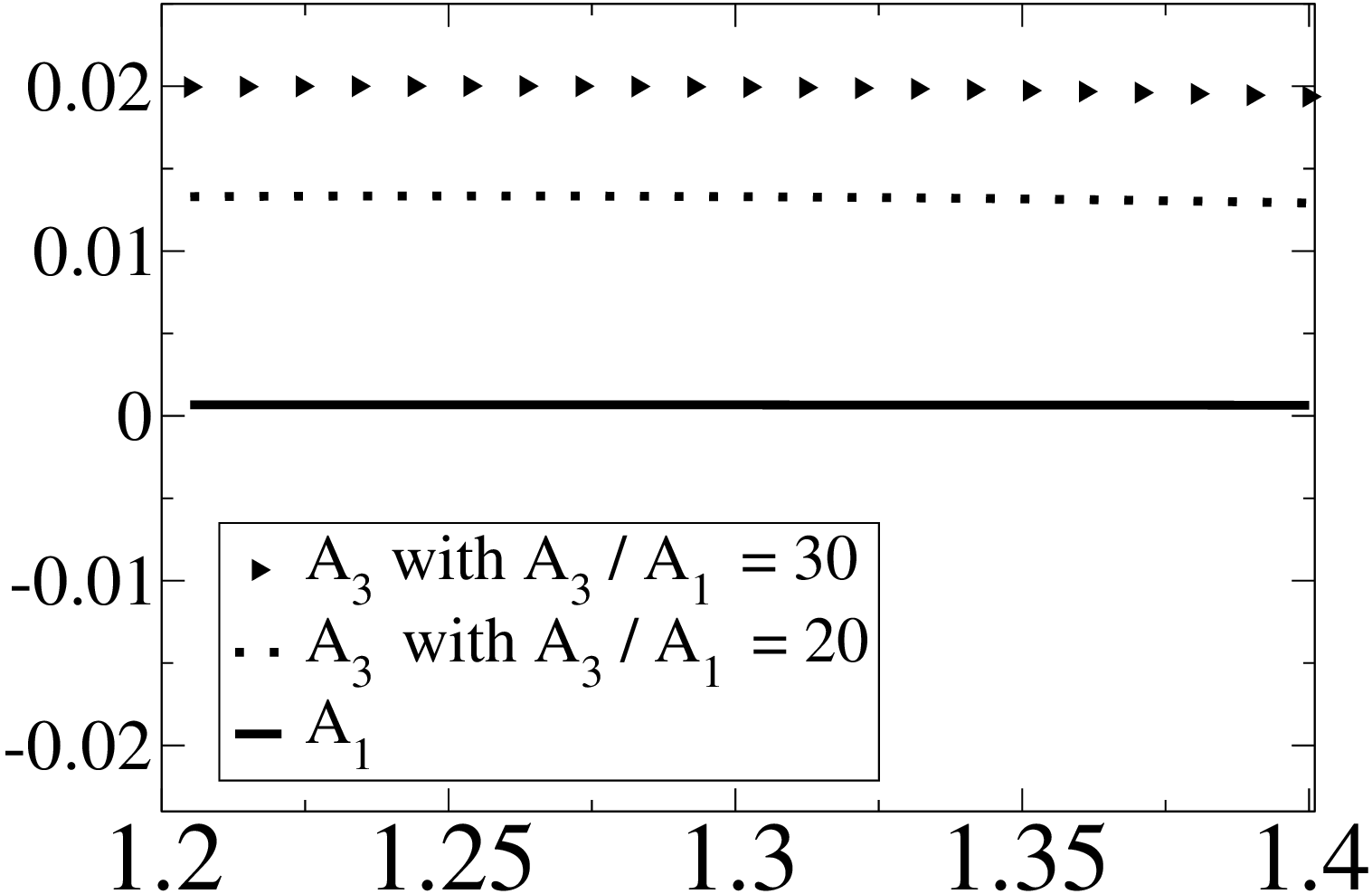}
\end{center}
\caption[]{%
Parameters vs $m[\pi(1300)]$ (GeV): $c_2$ 
(GeV$^2$) (top left), 
$d_2$ (GeV$^2$) (top right), $c_4$ (second row 
left), 
$e_3$ (GeV)(second row right), $\alpha_1$ 
and 
$\alpha_3$ (GeV)(third row left), $\beta_1$ 
and 
$\beta_3$ (GeV$^2$) (third row right) and $A_1$ 
and $A_3$ (GeV$^3$)
(last row).  (Note that $d_2$, $\alpha_1$, 
$\beta_1$ and $A_1$ do not depend on the choice 
of $A_3\over A_1$.)}
\label{Fig_par_vs_mpip}
\end{figure}

    It is interesting to compare the values of the
Lagrangian coefficients of the fully chiral invariant terms, 
$c_2,c_4^a,d_2,e_3^a$
as obtained here and displayed in Fig.\ref{Fig_par_vs_mpip}
 with the 
corresponding ones obtained
in the zero quark mass model: Fig.2 of \cite{2FJS07}.
These parameters are clearly substantially similar
to those in the zero quark mass case. This is a comforting
feature since it agrees with the expectation that
the light quark masses make only small changes in the QCD
dynamics.

    On the lowest two rows of Fig.\ref{Fig_par_vs_mpip} we display
the variations with $A_3/A_1$ and m[$\pi(1300)$] of the two-quark
condensates, the four-quark condensates and the quark mass type 
quantities $A_1$ and $A_3$.

\begin{figure}
\begin{center}
\vskip 1cm
%-----------------------------------------
\epsfxsize = 6cm
 \epsfbox{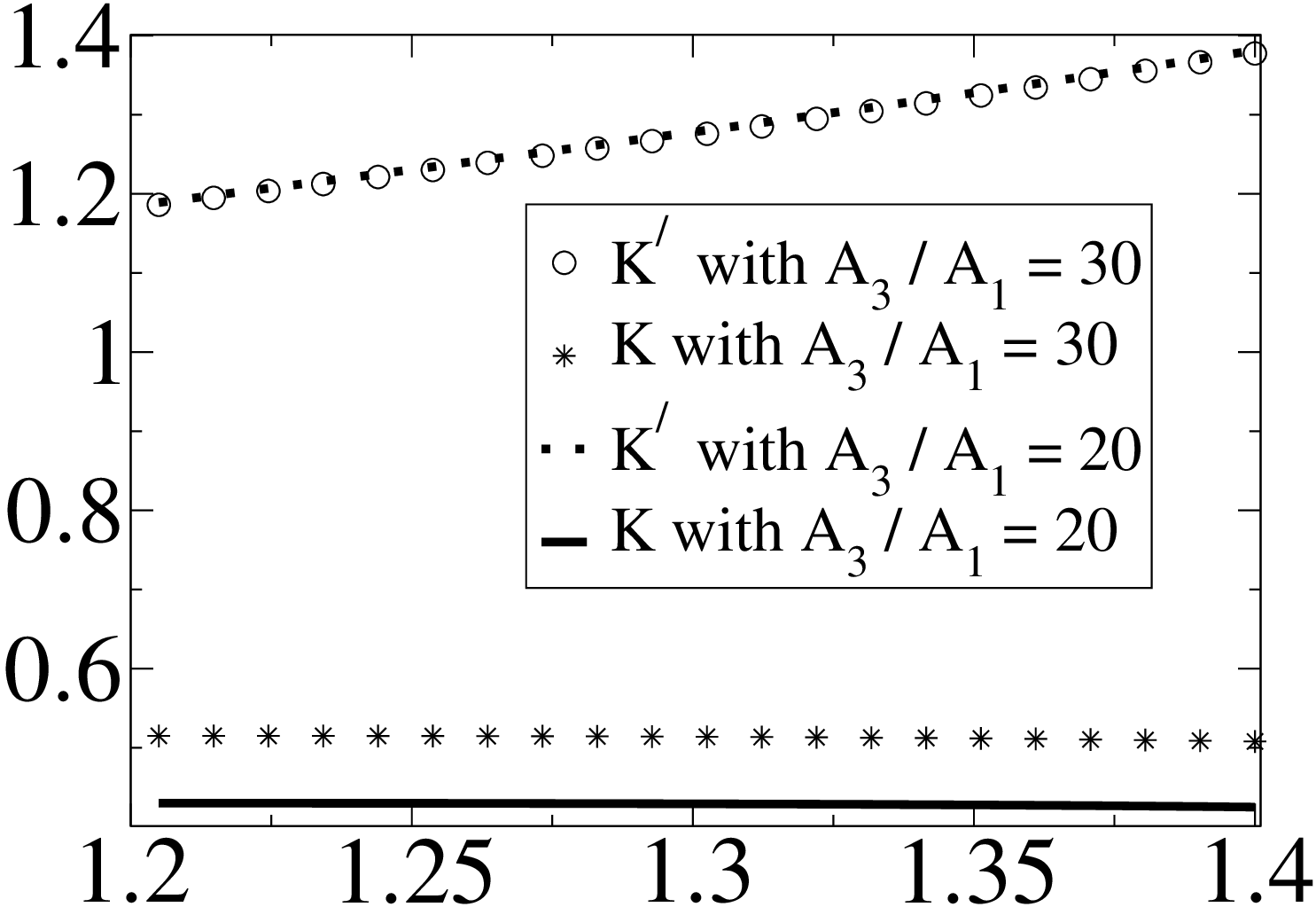}
\hskip 1cm
\epsfxsize = 6cm
 \epsfbox{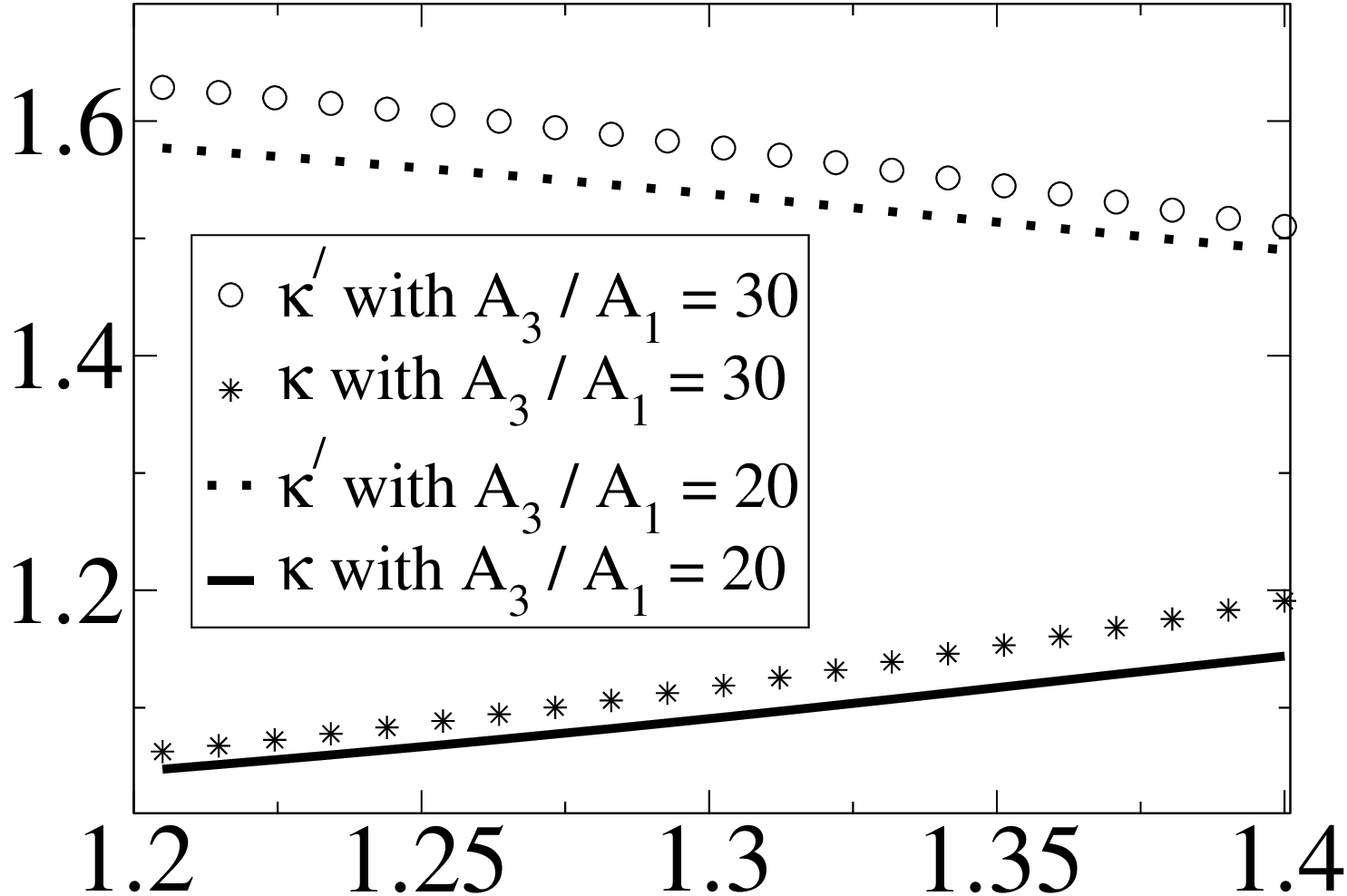}
\vskip 1cm
%-----------------------------------------
\epsfxsize = 6cm
 \epsfbox{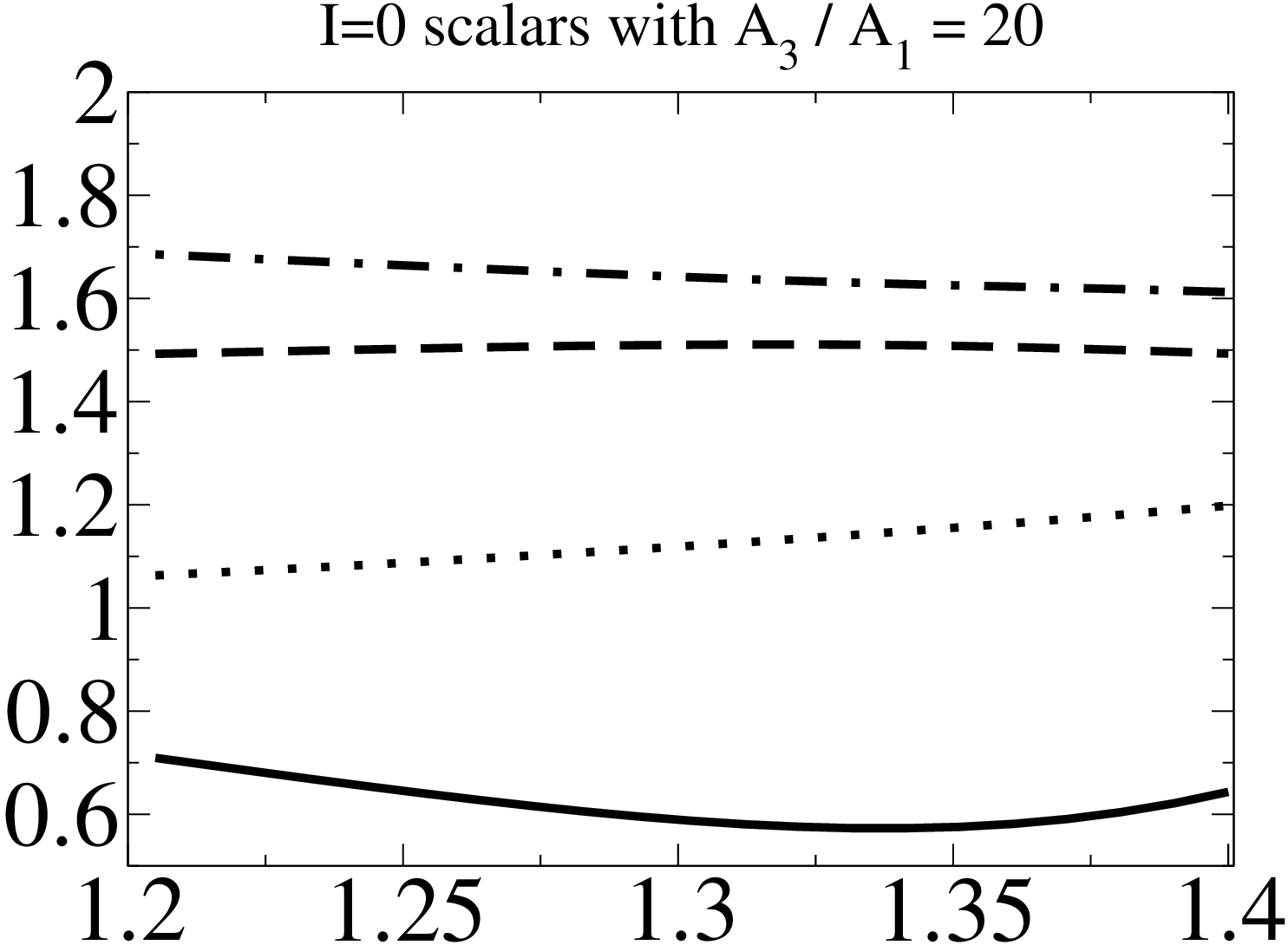}
\hskip 1cm
\epsfxsize = 6cm
 \epsfbox{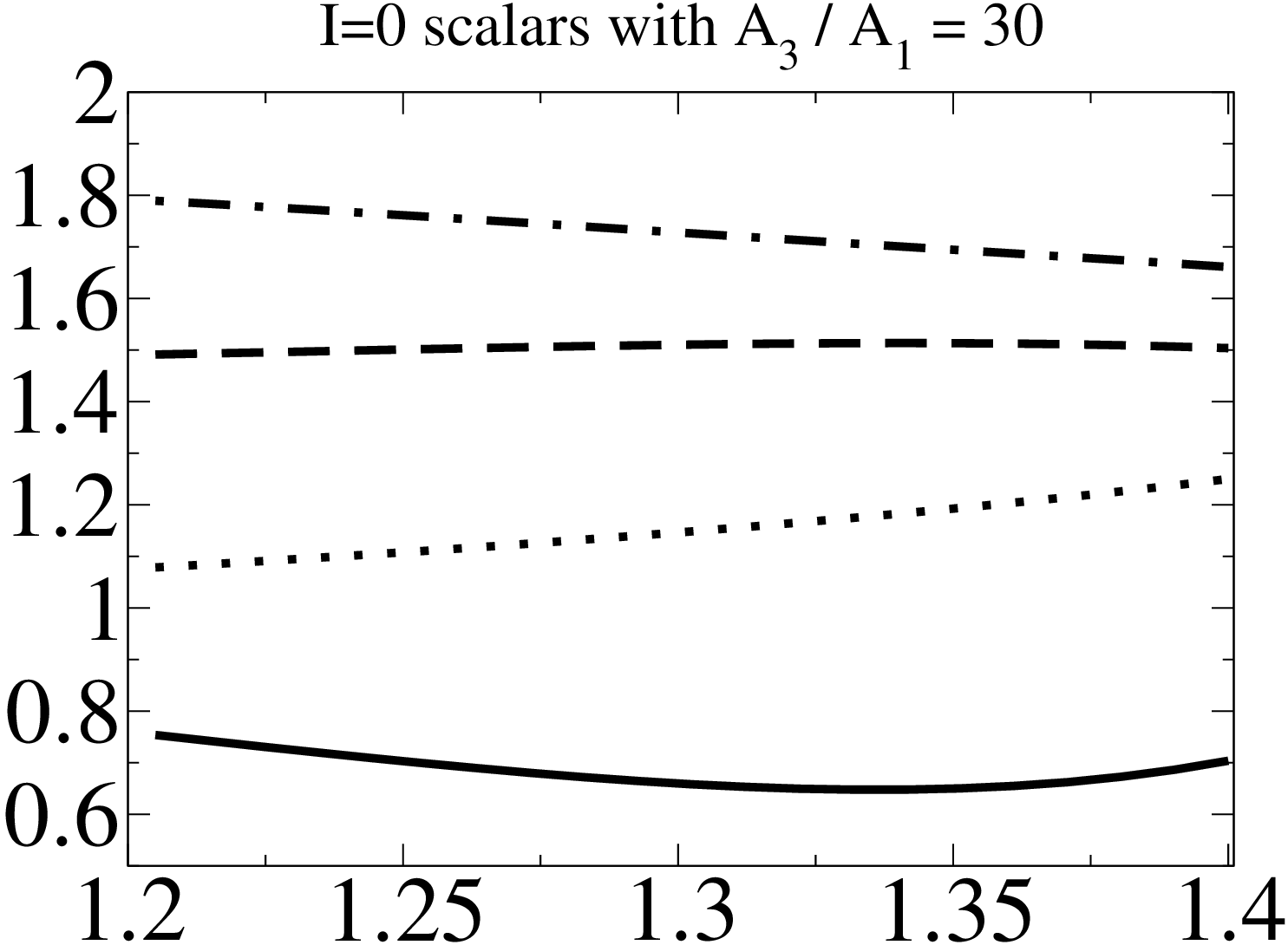}
\end{center}
\caption[]{%
Predicted masses (GeV) vs $m[\pi(1300)]$ (GeV): 
kaon system 
(top left), 
kappa system (top right), isoscalar salars 
(last row).
}
\label{Fig_masses_vs_mpip}
\end{figure}

    In Fig.\ref{Fig_masses_vs_mpip} we give the predictions
for the masses of the strange mesons and the I=0 scalars,
showing the variations due to choosing different values of
m[$\pi(1300)$] and $A_3/A_1$. It is seen that the mass of the
ordinary kaon is better fit for $A_3/A_1$ closer to 30 
than to 20. The predicted mass of the ``excited" kaon is 
roughly similar to the (mentioned, but not established)
candidate K(1460).

    The upper, right graph of Fig.\ref{Fig_masses_vs_mpip}
displays the two predicted kappa-type particle masses.
These are roughly consistent with the (discussed
in \cite{RPP} but  
not established)$K_0^*$(800) and the $K_0^*$(1430).
It should be mentioned that the masses 
obtained in this paper are being considered as  ``bare"
ones, subject to non-trivial
``renormalization" due to effects which
provide unitarity corrections to the scattering amplitudes
in which these particles appear as poles.
 This is illustrated
for the single-M linear sigma model case in 
\cite{{BFMNS01}}. A similar remark
 applies to the particle
widths obtained from the third derivatives of the 
potential.

    The graphs in the bottom row of
 Fig.\ref{Fig_masses_vs_mpip} display the predicted
masses of the four I=0 scalars. Clearly the lighter
two can be identified with the $f_0$(600)[sigma]
and the $f_0$(980). Note that the very light sigma
is an inevitable consequence of the present model
and was not put in by hand. There are three
 candidates,$f_0$(1370),$f_0$(1500)
and $f_0$(1710)[one of which may be a glueball],
for the heavier two particles.

\begin{figure}
\begin{center}
\vskip 1cm
%-----------------------------------------
\epsfxsize = 6cm
 \epsfbox{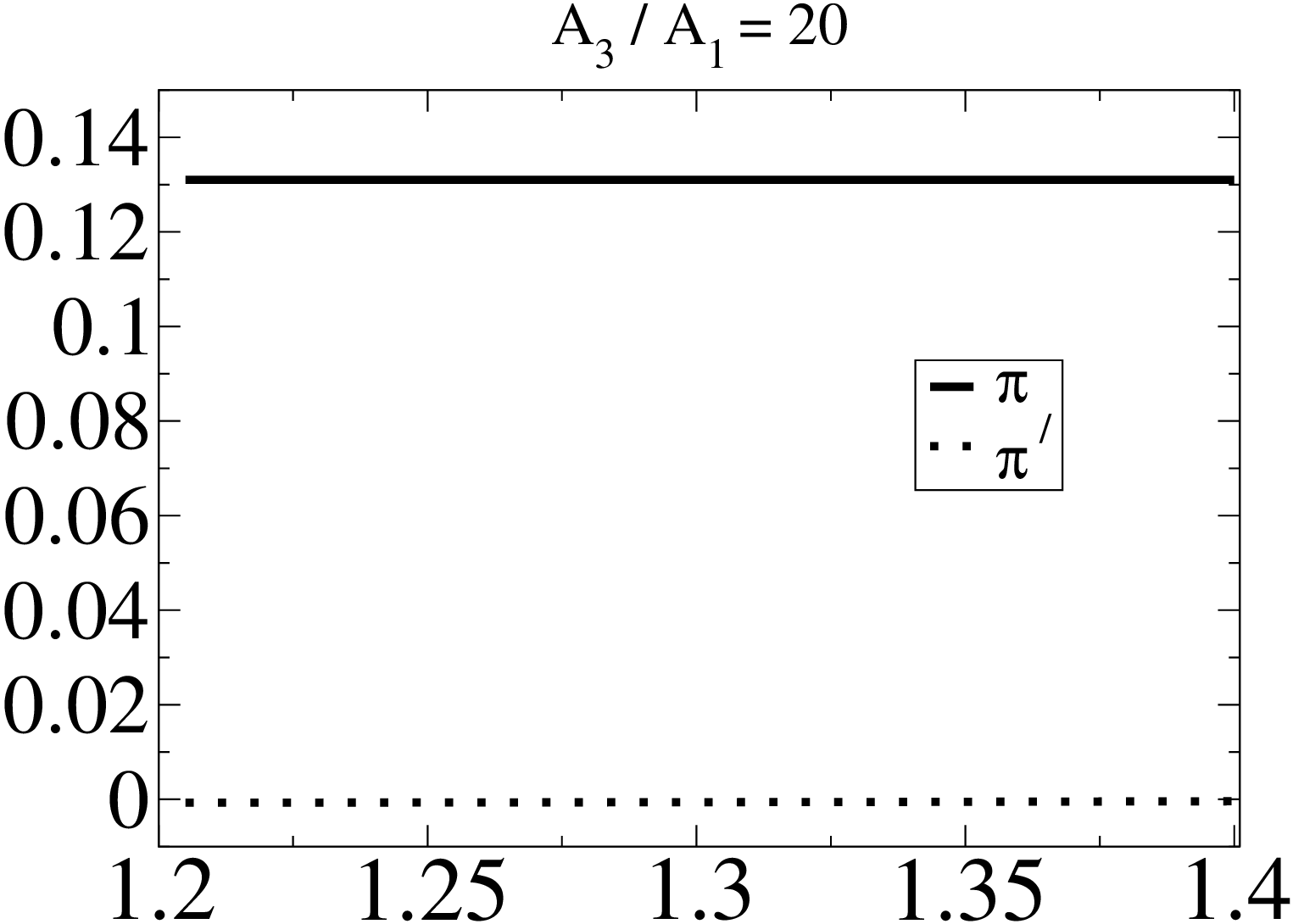}
\hskip 1cm
\epsfxsize = 6cm
 \epsfbox{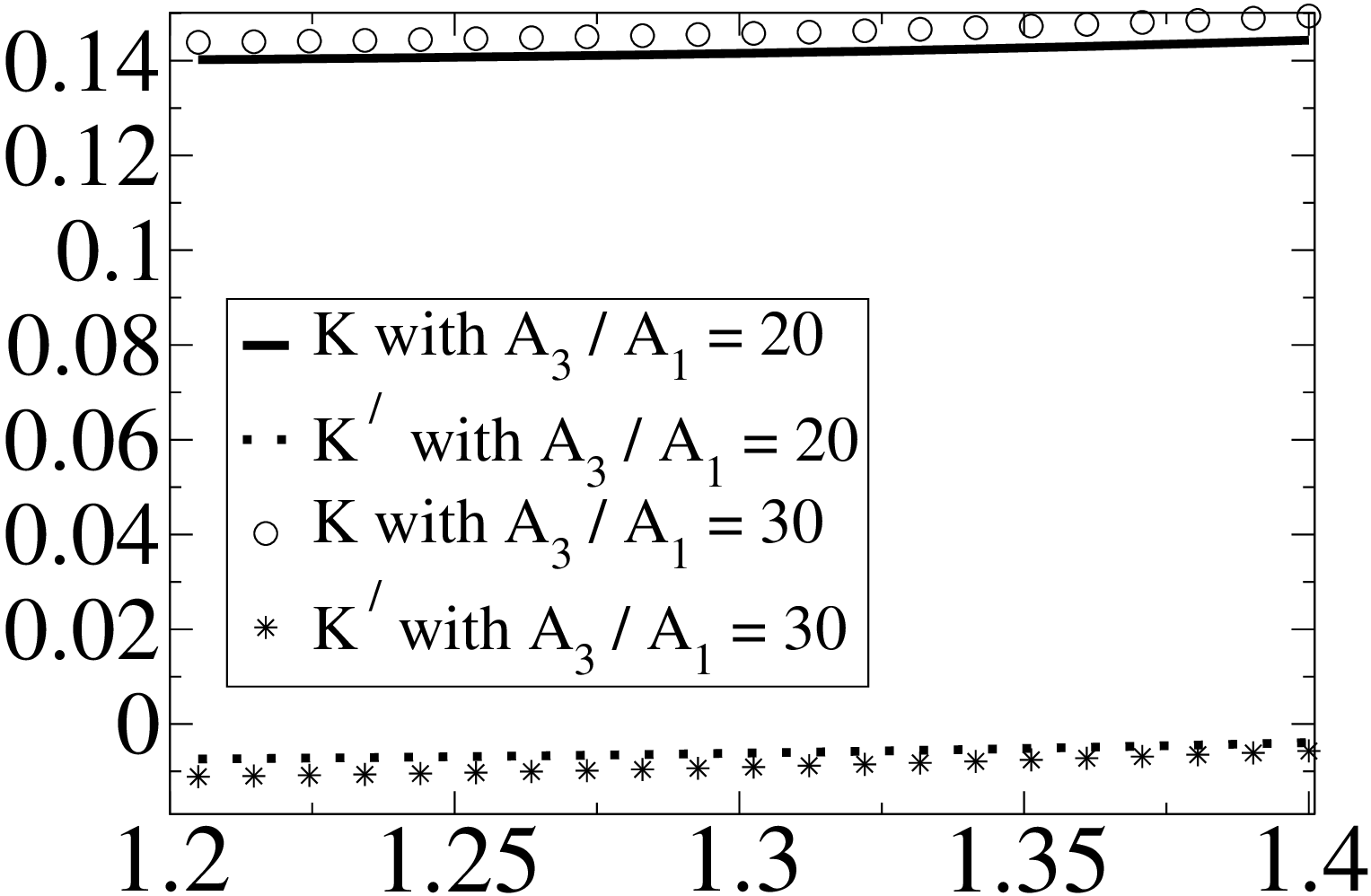}
\vskip 1cm
%-----------------------------------------
\epsfxsize = 6cm
 \epsfbox{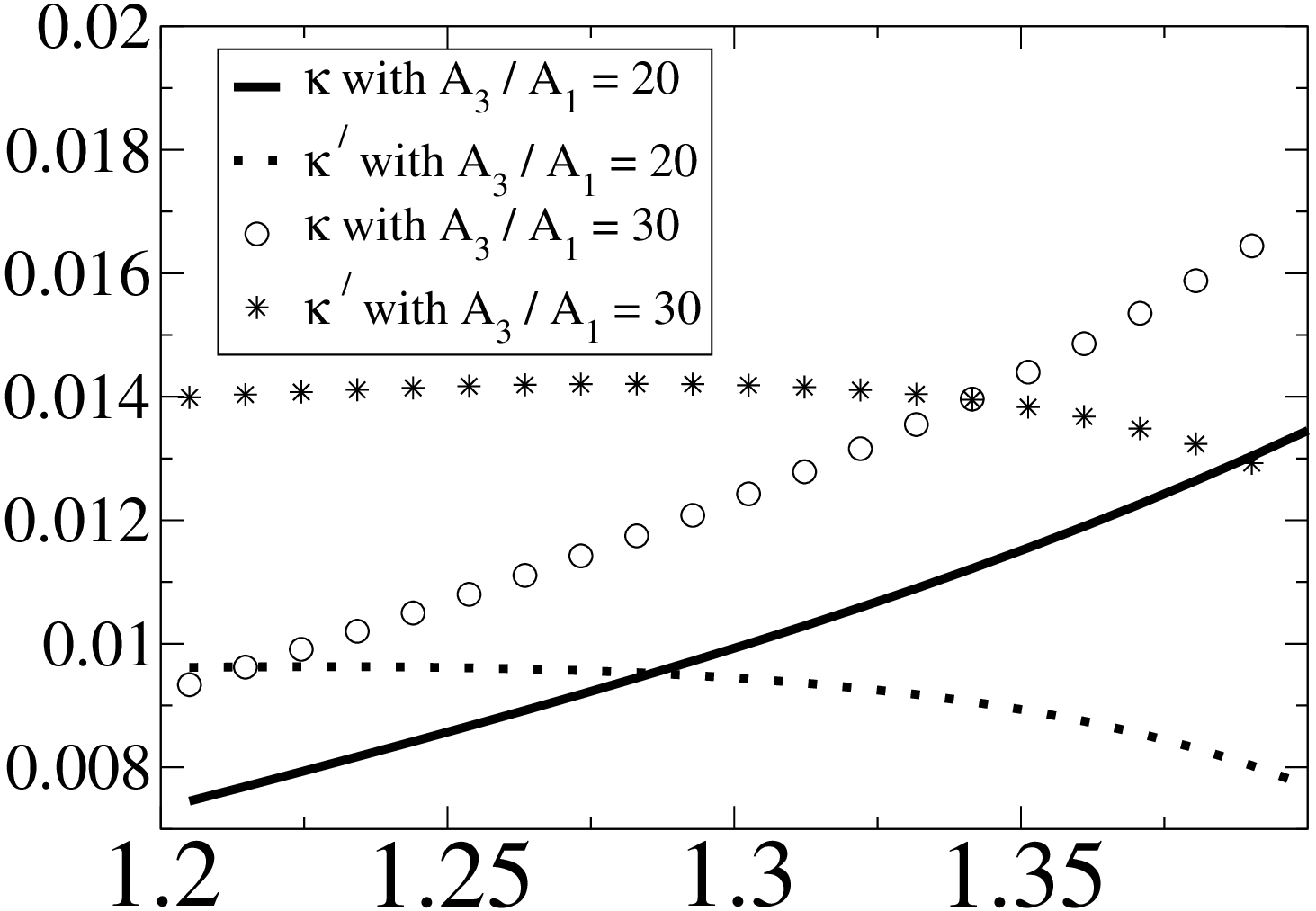}
\end{center}
\caption[]{%
Predicted decay constants (GeV) vs 
$m[\pi(1300)]$ (GeV): 
pion system 
(top left), 
kaon (top right), kappa system 
(last row).   The result for pion system is 
insensitive to $A_3/A_1$.
}
\label{Fig_Fkpi_vs_mpip}
\end{figure}

    In Fig.\ref{Fig_Fkpi_vs_mpip} the predicted
decay constants (i.e. the coefficients of the single 
particle terms in the appropriate axial vector or 
vector Noether currents) for the $\pi'$, the K, 
the K', the $\kappa$ and the $\kappa'$ are shown
with their parameter dependences.
Note that $F_{\pi'}$ is very small but not exactly
zero. This causes, as discussed in \cite{3FJS07}, a 
very small violation  of the "partially conserved"
axial vector current ansatz in this kind of mixing
model.

\begin{figure}
\begin{center}
\vskip 1cm
%-----------------------------------------
\epsfxsize = 6cm
 \epsfbox{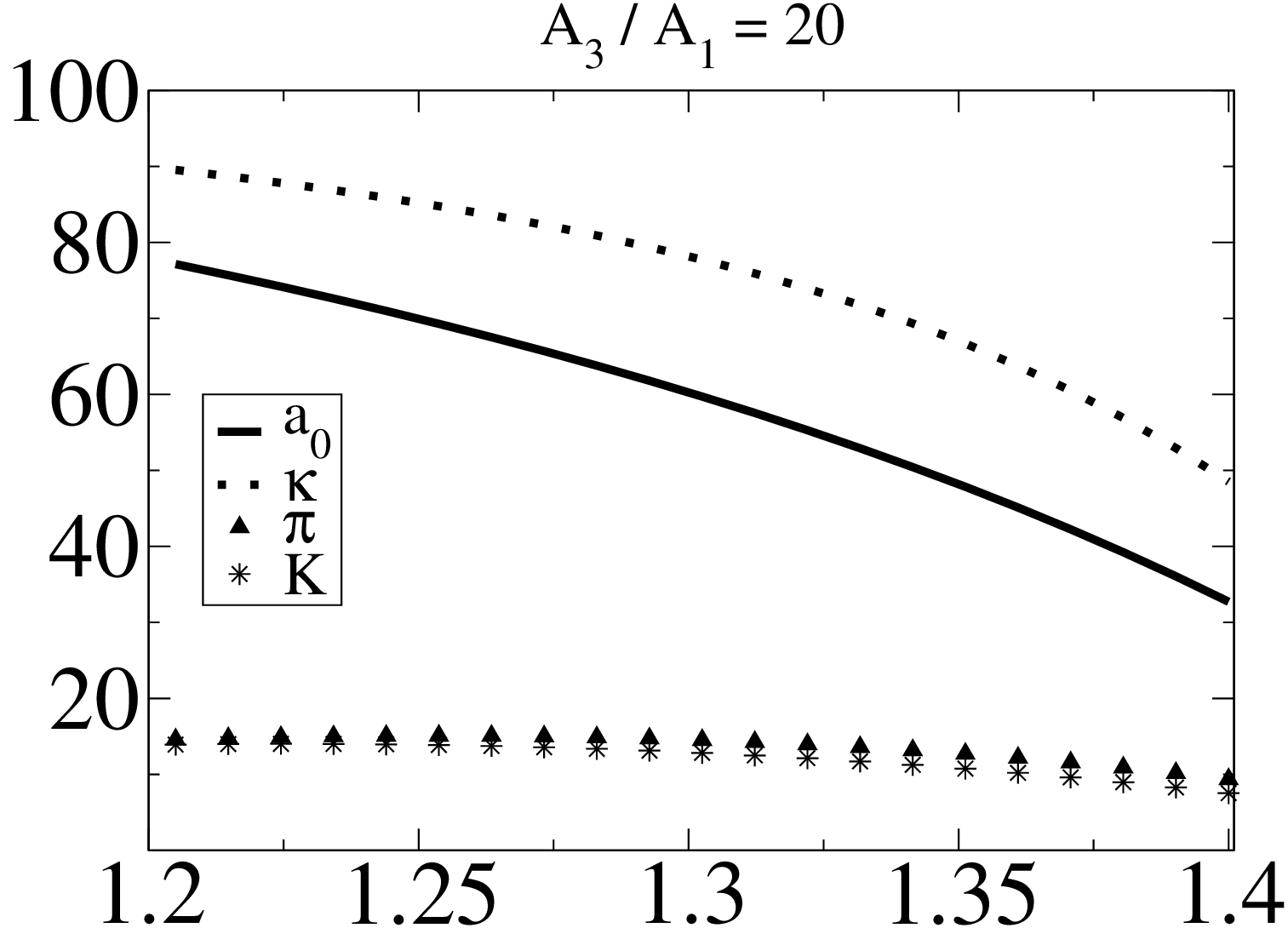}
\hskip 1cm
\epsfxsize = 6cm
 \epsfbox{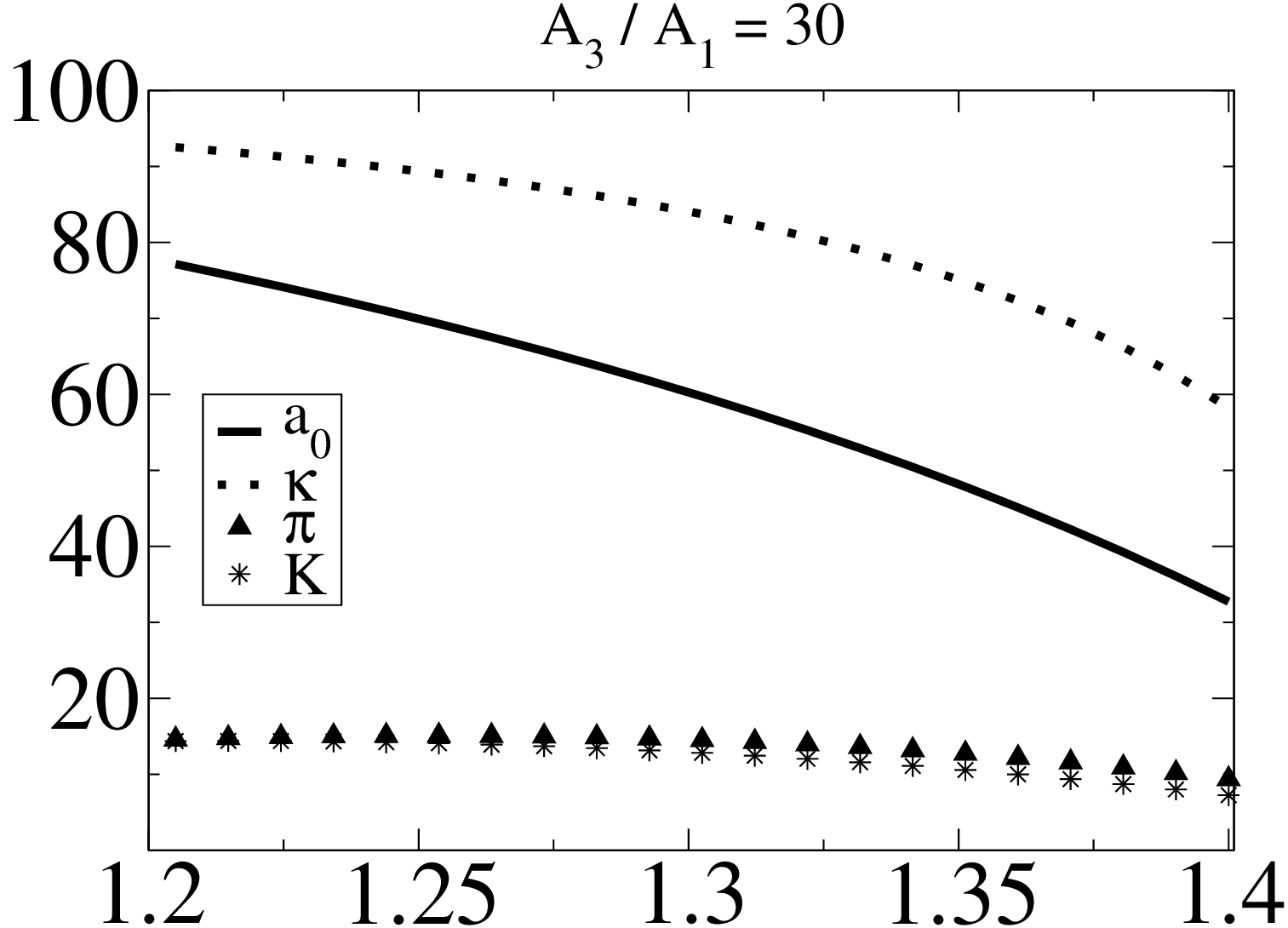}
\end{center}
\caption[]{%
Predicted percentage of four quark contents  vs 
$m[\pi(1300)]$ (GeV).
}
\label{Fig_4q_vs_mpip}
\end{figure}

    One of our chief concerns here is the 
percentage ``content" of ``two quarks" (i.e.
the $q{\bar q}$ content) vs. the percentage 
content of ``four quarks" predicted to be
in each particle state. In Fig.\ref{Fig_4q_vs_mpip}
these percentages are shown for the particles 
having non-zero isospin quantum number.
For comparison with the SU(3) symmetric 
cases considered in \cite{2FJS07}
and \cite{3FJS07}, note that the typical
value of $m[\pi(1300)]$ considered there
was about 1215 MeV. Then we see that the
ordinary pion has, as before, about a 
15 percent 4-quark content (and 85 percent
2-quark content). The ``excited" pion,
$\pi(1300)$ is predicted to have an 
85 percent 4-quark content in this picture.
Similarly, we see that the ordinary kaon
is predicted to have about a 14 percent 
4-quark content while the ``excited" kaon has
about an 86 percent 4-quark content. Notably,
 as before, the situation is
drastically different for the 
scalar states. We read off that the 
lighter I=1 scalar, $a_0$(980) has 
about a 76 percent 4-quark content and
the lighter kappa has about a 92 percent
4-quark content.  

\begin{figure}
\begin{center}
\vskip 1cm
%-----------------------------------------
\epsfxsize = 6cm
 \epsfbox{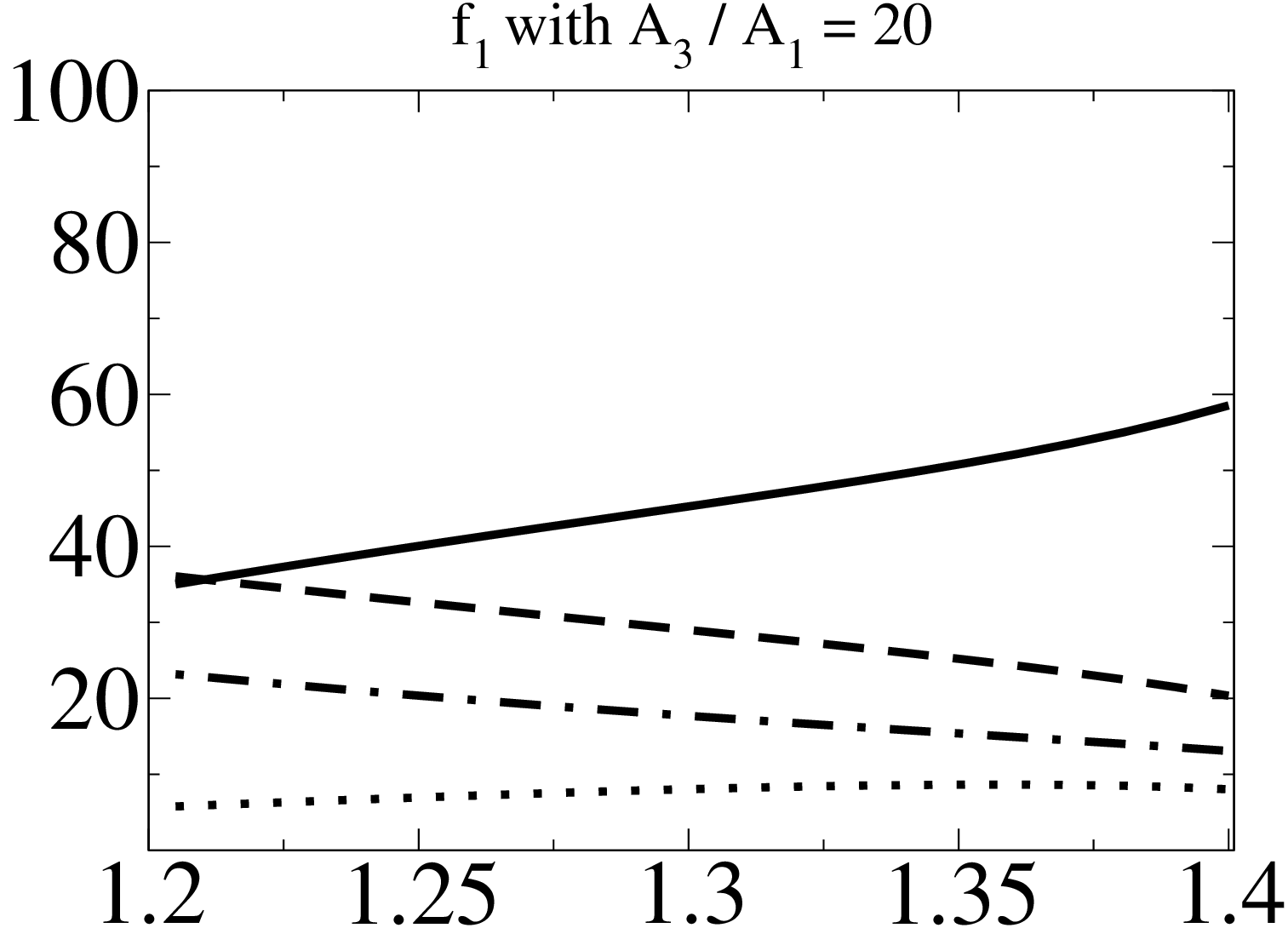}
\hskip 1cm
\epsfxsize = 6cm
 \epsfbox{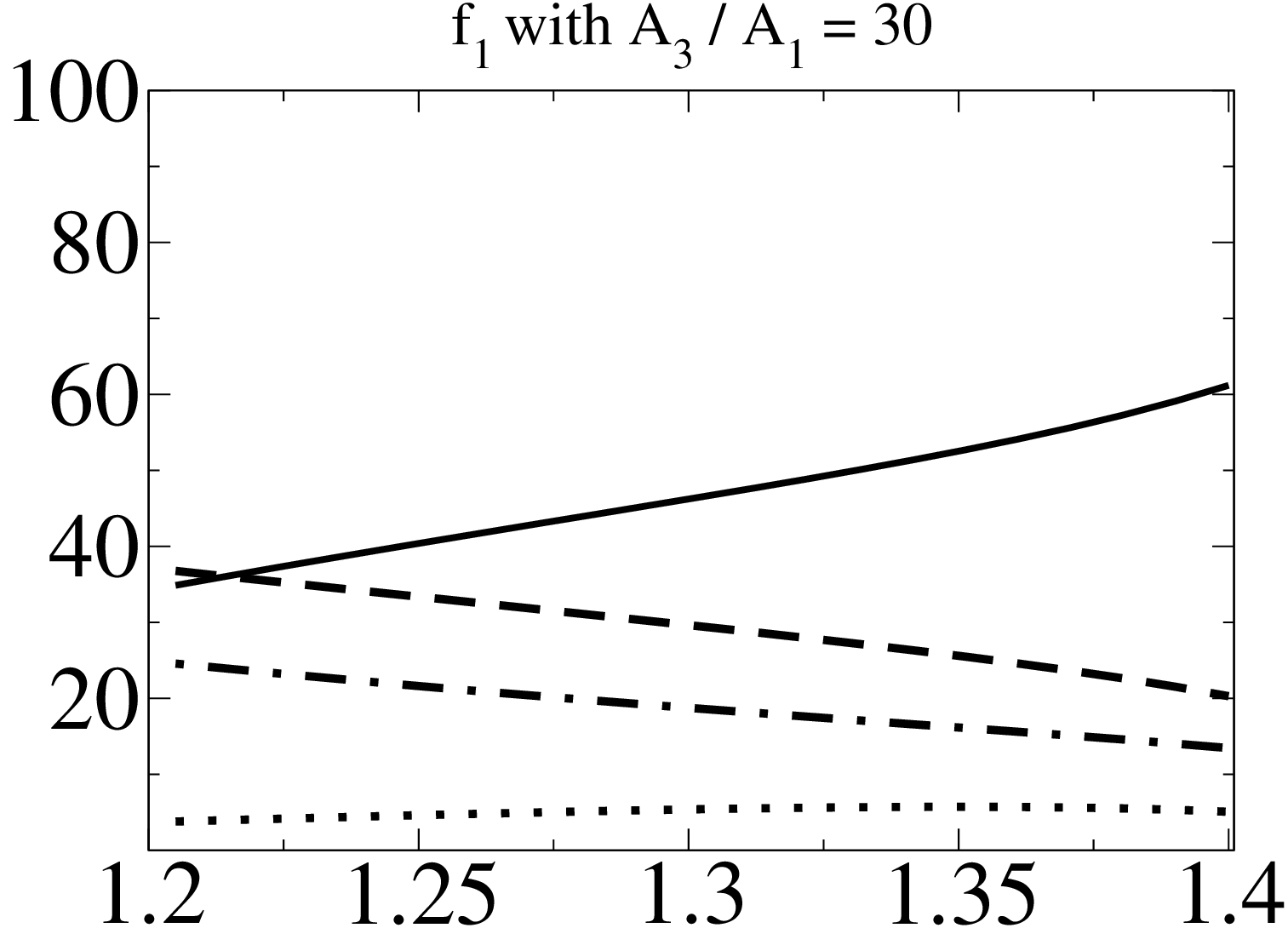}
\vskip 1cm
%-----------------------------------------
\epsfxsize = 6cm
 \epsfbox{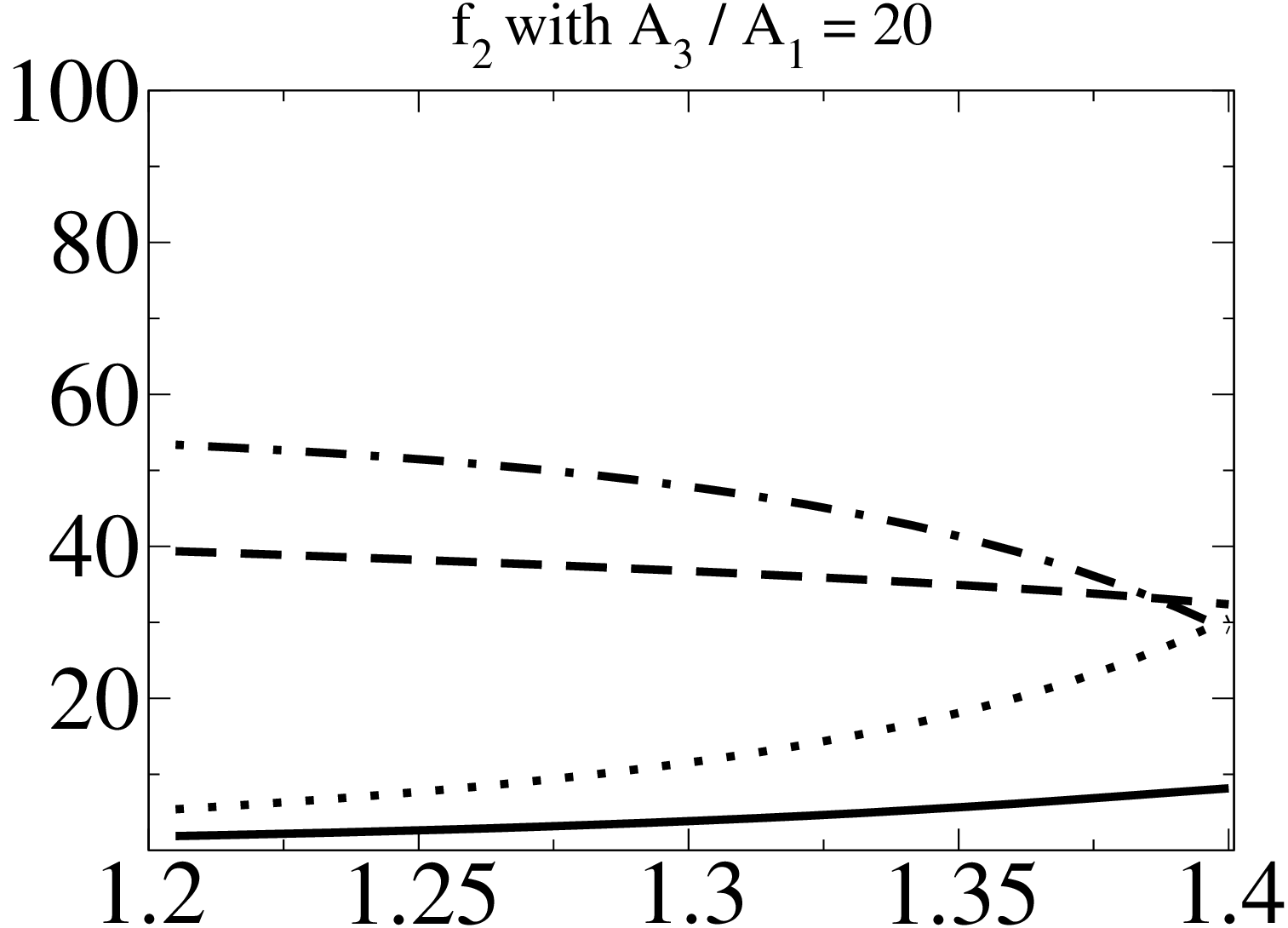}
\hskip 1cm
\epsfxsize = 6cm
 \epsfbox{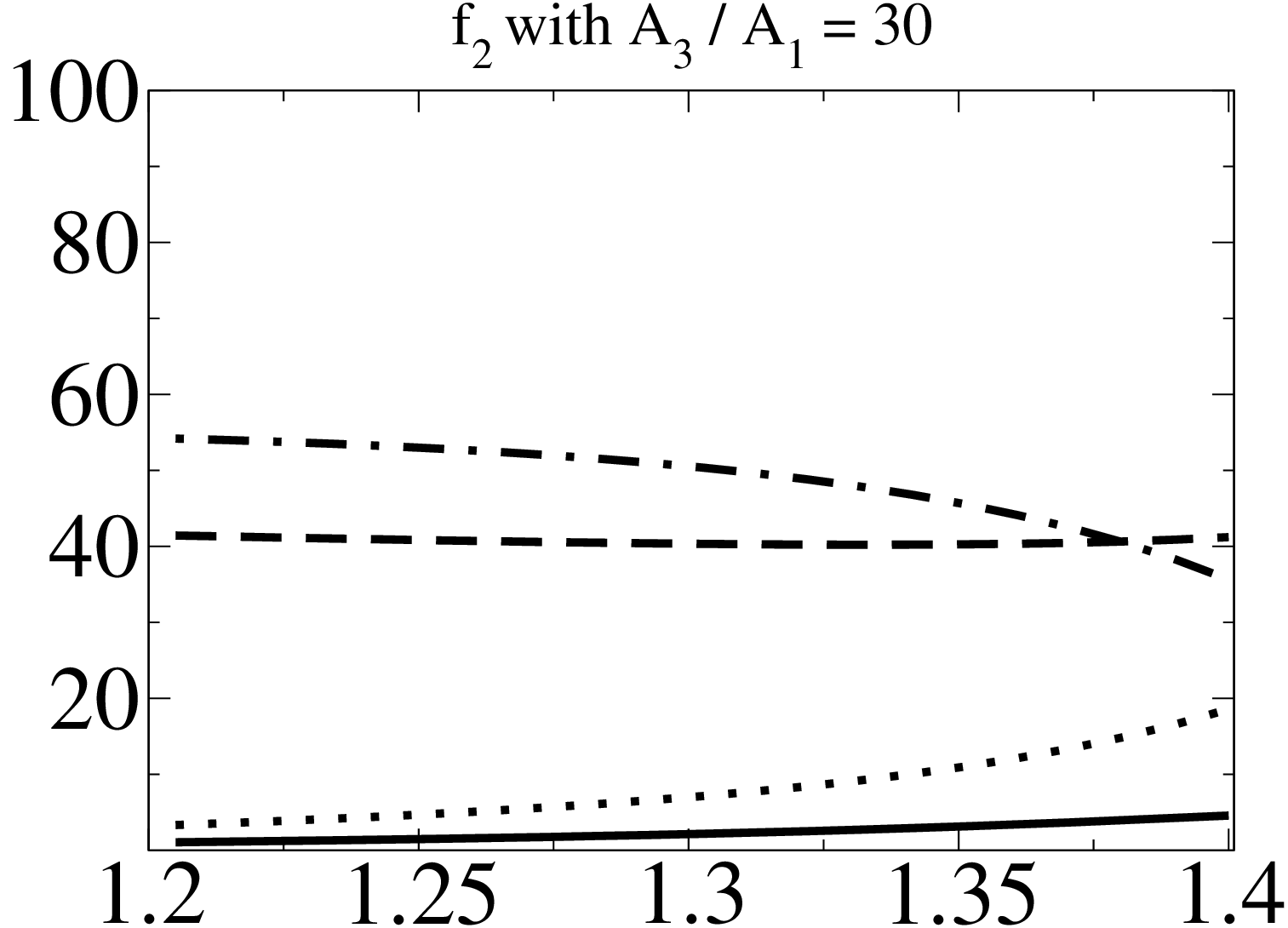}
%-----------------------------------------
\vskip 1cm
%-----------------------------------------
\epsfxsize = 6cm
 \epsfbox{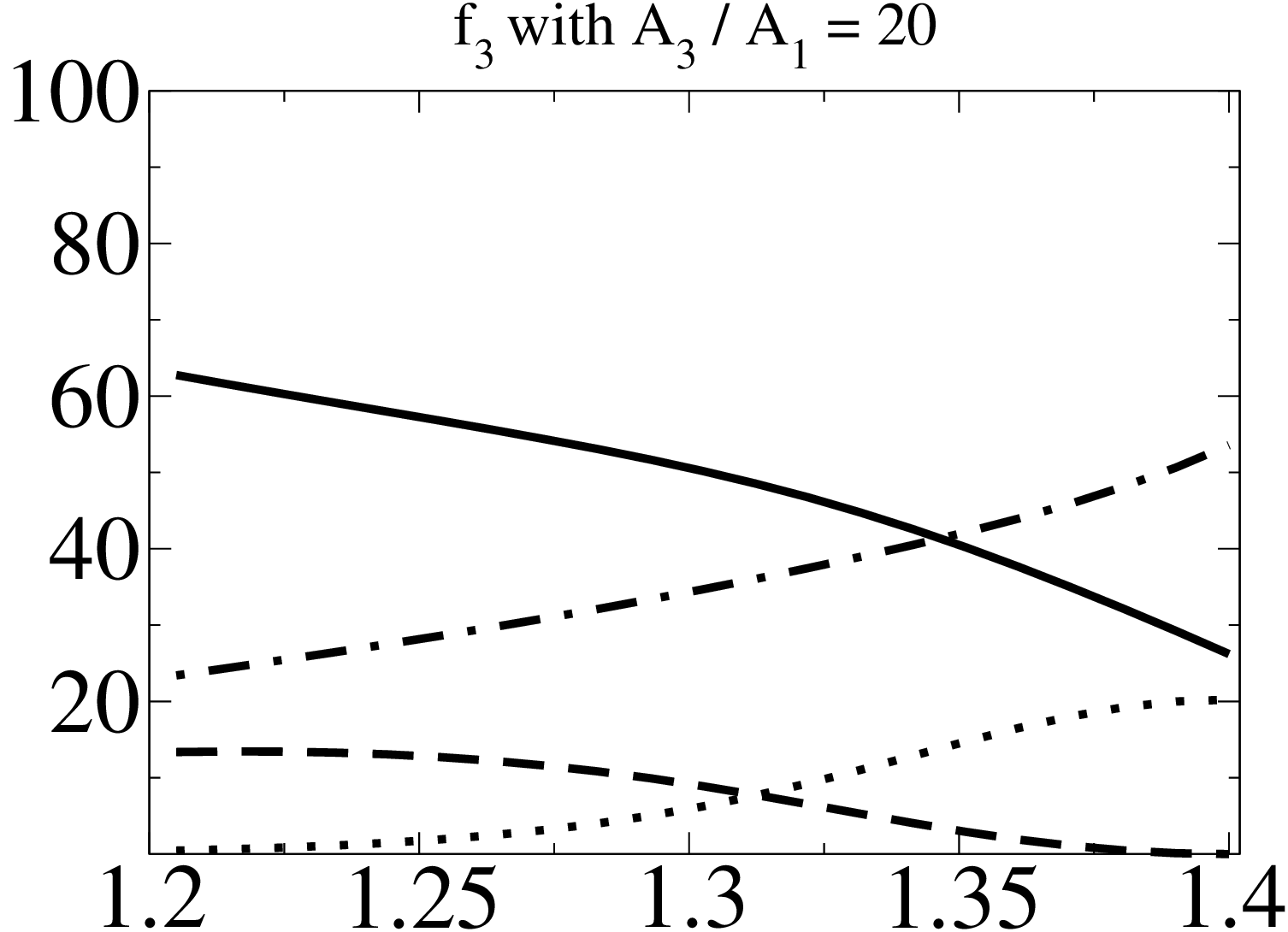}
\hskip 1cm
\epsfxsize = 6cm
 \epsfbox{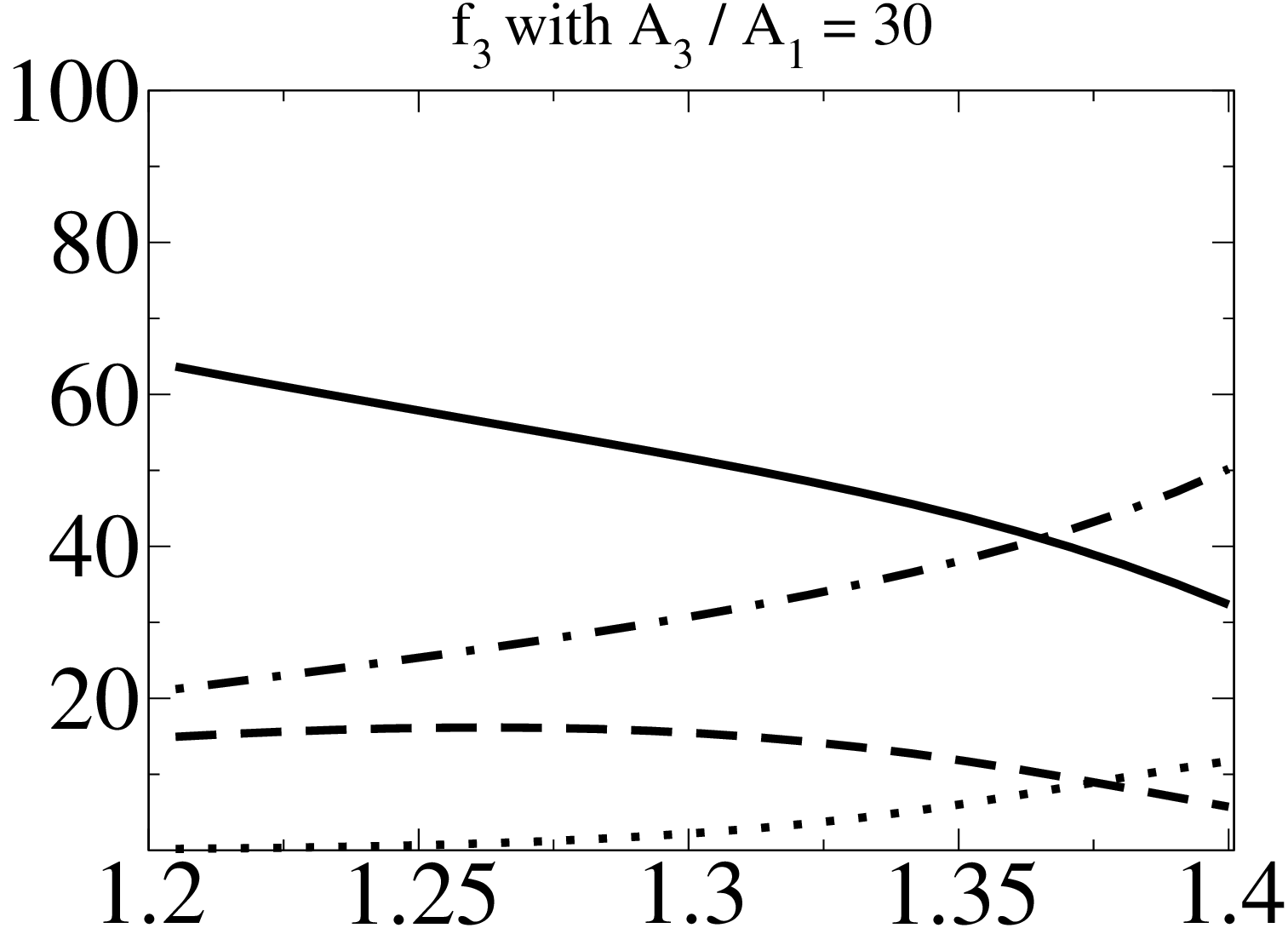}
\vskip 1cm
%-----------------------------------------
\epsfxsize = 6cm
 \epsfbox{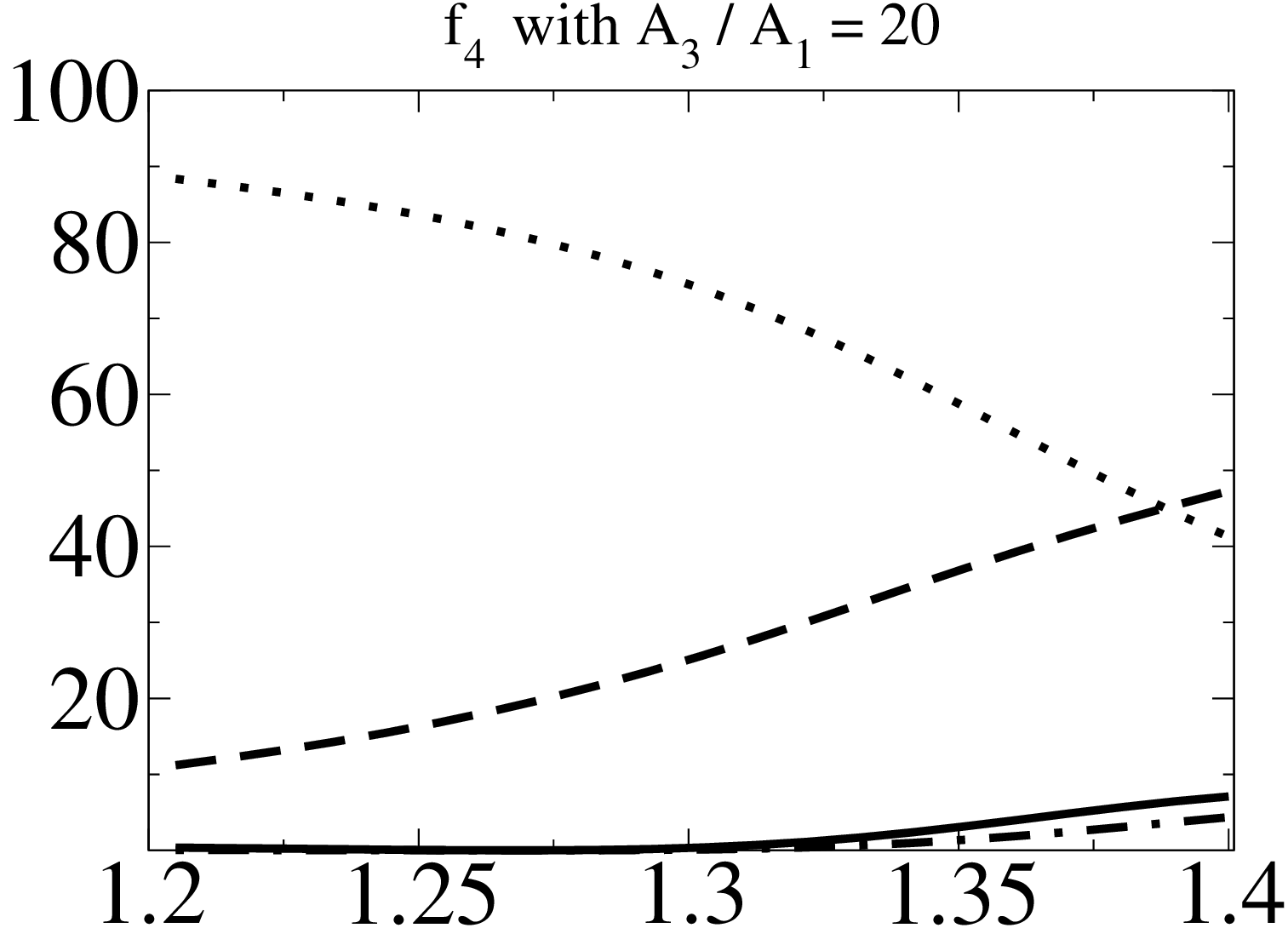}
\hskip 1cm
\epsfxsize = 6cm
 \epsfbox{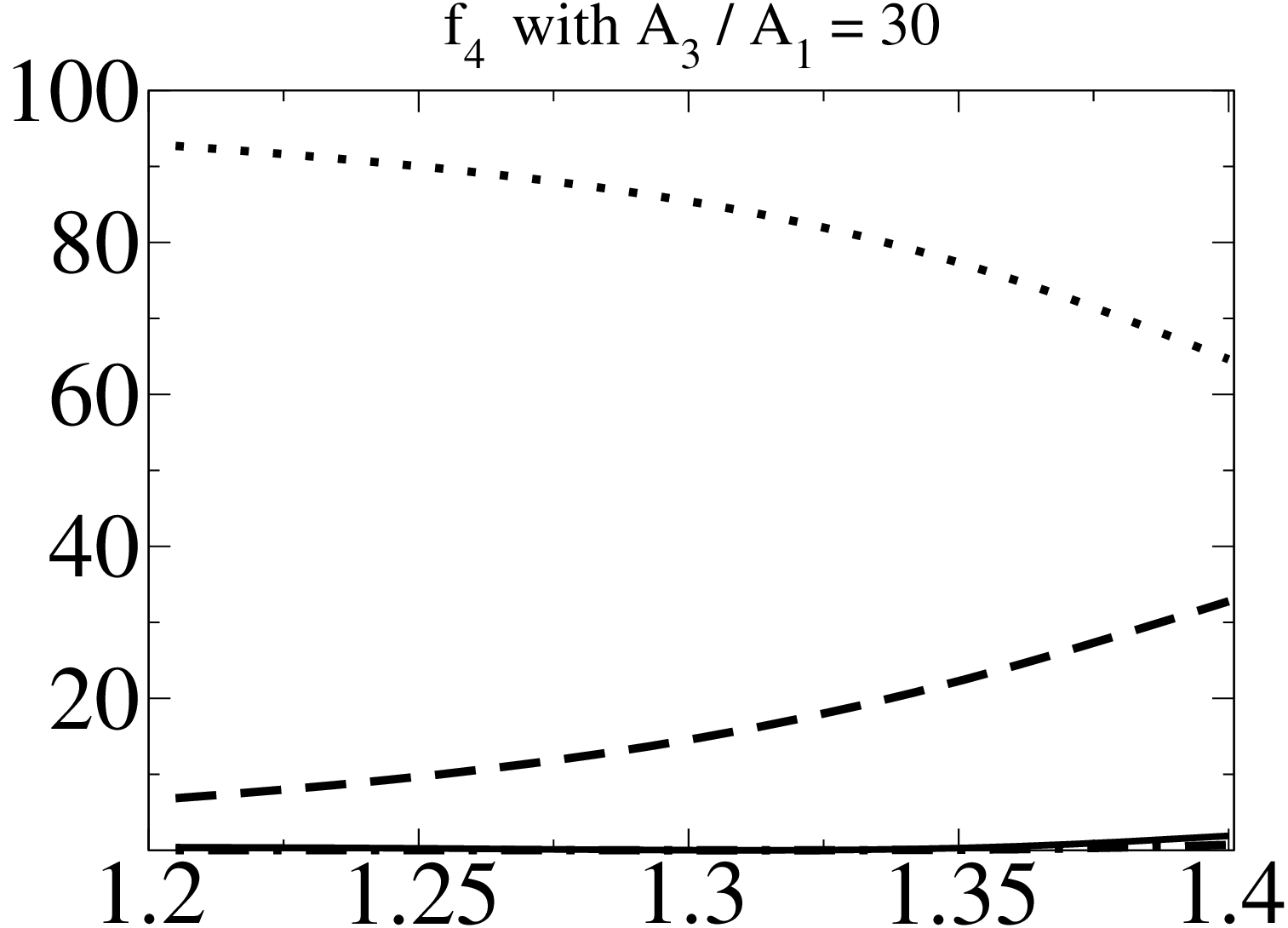}
%-----------------------------------------
\end{center}
\caption[]{%
Predicted quark contents of isoscalar scalar 
mesons vs	   
$m[\pi(1300)]$ (GeV).   The components 
[described in Eq. (\ref{fourbasis})] are:
$f_a$ (solid line), 
$f_b$ (dotted line), 
$f_c$ (dashed line) and  
$f_d$ (dotted-dashed line). 
}
\label{Fig_fcomps_vs_mpip}
\end{figure}

   The more complicated situation of the
four mixing I=0 scalars is described in
Fig.\ref{Fig_fcomps_vs_mpip}. To read this, first 
note that the four physical states are labelled
$f_1,f_2,f_3,f_4$ in order of increasing mass.
Each is a linear combination of the basis states
$f_a,f_b,f_c,f_d$ given
 in Eq.(\ref{fourbasis}). $f_a$ and $f_b$
are 2-quark type while $f_c$ and $f_d$
are 4-quark type. Thus the lowest lying 
I=0 scalar, $f_1\equiv\sigma$, taking 
$A_3/A_1$=30 and $m[\pi(1300)]$ = 1215 MeV,
has percentages in each basis state of,
\begin{equation}
0.36,\,\,  0.04,\,\, 0.36,\,\, 0.24.
\label{sigmadecomp}
\end{equation}
Altogether the sigma is about 40 percent 2-quark
and 60 percent 4-quark in this case. This is
similar to the roughly 50-50 split we found in the 
SU(3) symmetric case (See Fig.2 of\cite{3FJS07}).
$f_2$, the next heaviest I=0 state is about
95 percent 4-quark and 5 percent 2-quark.
The two heaviest I=0 scalars both have 
 majority two quark nature: $f_3$ is
 read to be about 63 percent 2-quark
while $f_4$ is read to be about 93 percent
2-quark.  

\section{I=0 pseudoscalars}

\begin{figure}
\begin{center}
\vskip 1cm
%-----------------------------------------
\epsfxsize = 6cm
\epsfbox{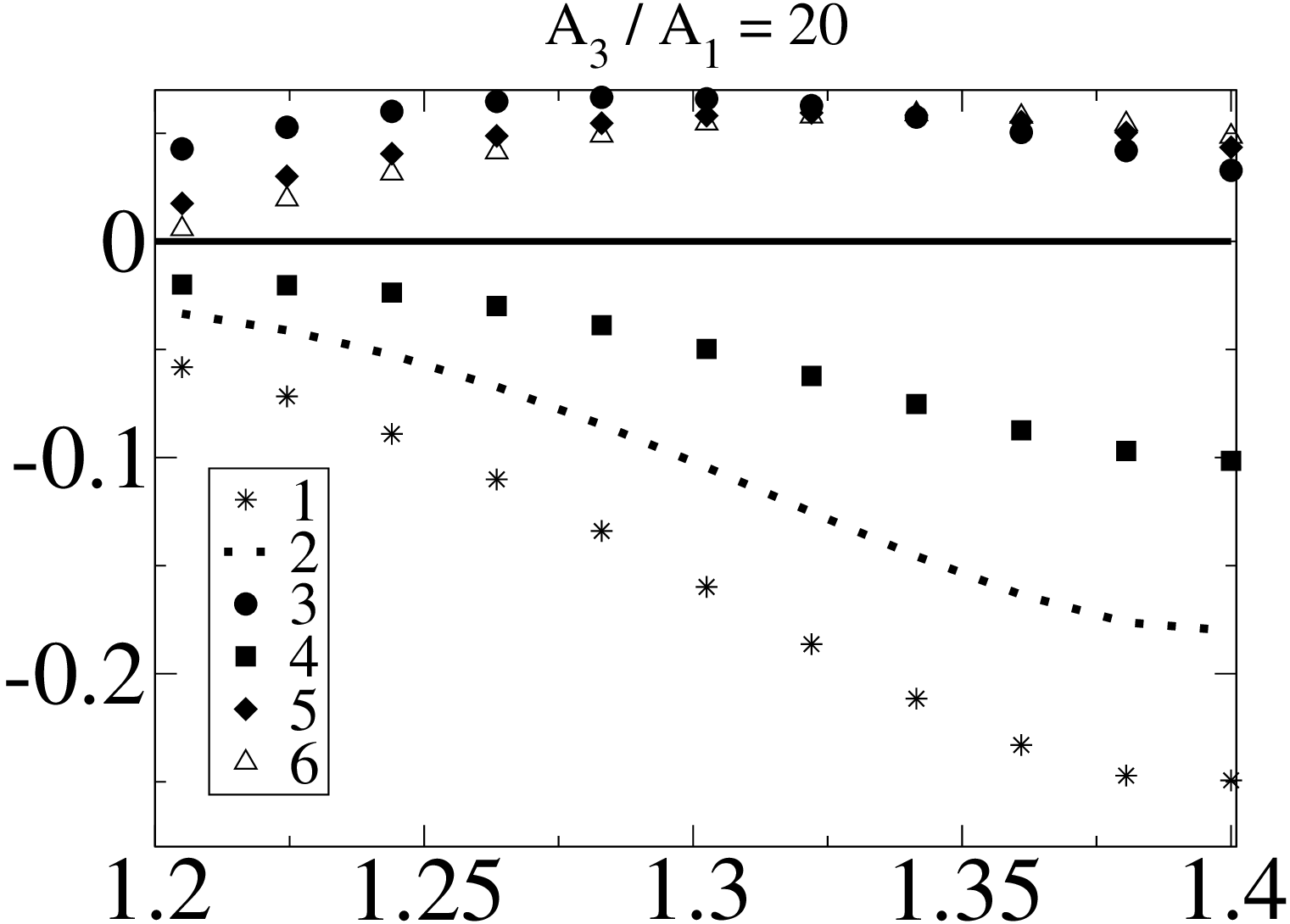}
%-----------------------------------------
\epsfxsize = 6cm
\epsfbox{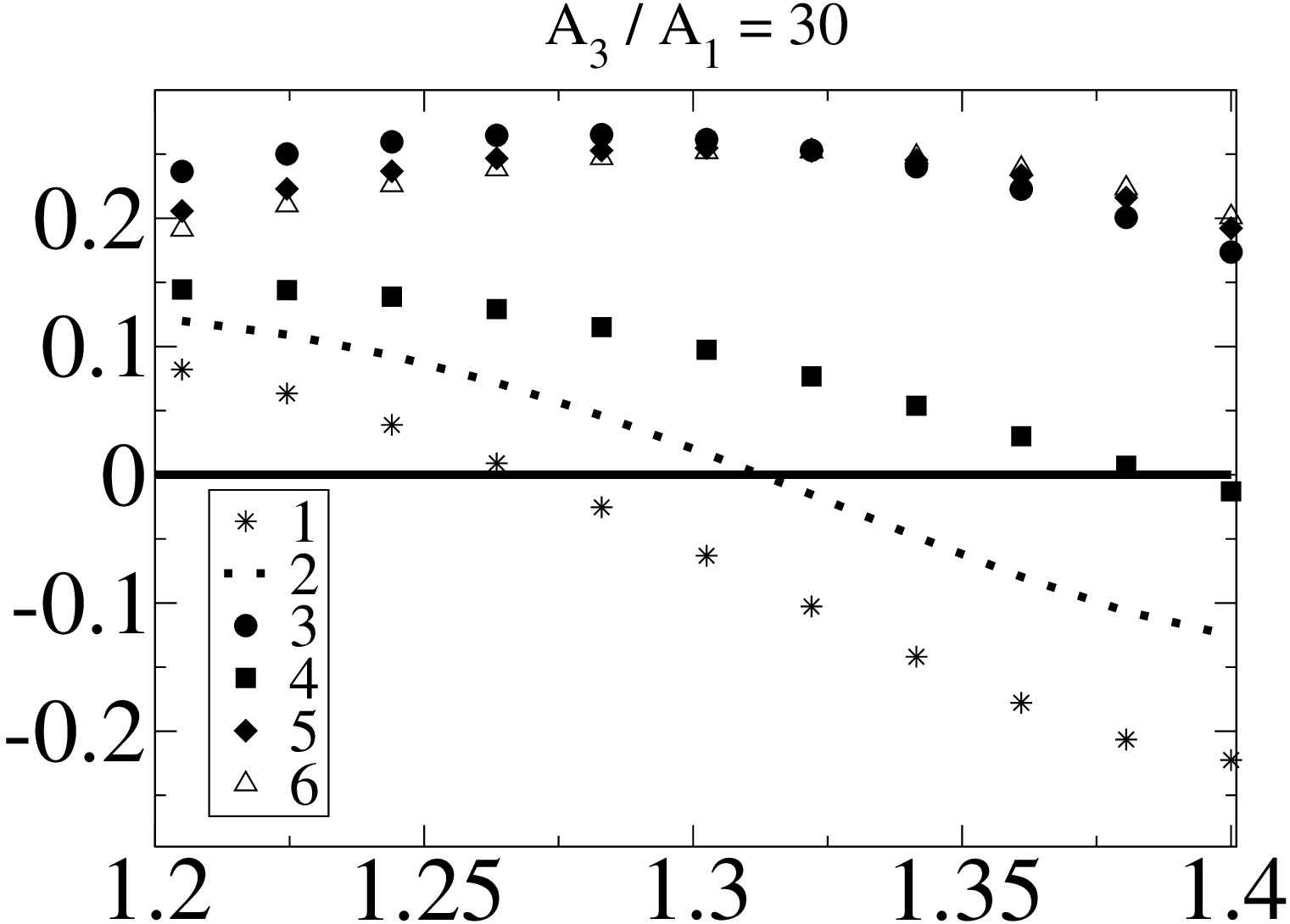}
\hskip 1cm
%-----------------------------------------
\end{center}
\caption[]{%
Discriminant vs	   
$m[\pi(1300)]$ (GeV).
}
\label{Fig_dscm}
\end{figure}

The elements of the squared mass matrix for I=0
pseudoscalars (given in Appendix B)  are quite
complicated and as a result it is convenient to
treat this sector separately.
As previously discussed,   in this case the 
squared mass matrix 
contains two 
additional 
parameters ($c_3$ and $\gamma_1$) that only 
contribute to the properties of the I=0 
pseudoscalars.   It is therefore not possible 
to trade these two parameters 
with four  experimental $\eta$ masses.
Instead, we determine 
$c_3$ and $\gamma_1$ by 
fitting the trace 
and 
the determinant of the squared mass matrix to 
their 
corresponding experimental values, i.e. we 
solve for $c_3$ and $\gamma_1$ from the 
following equations:
\begin{eqnarray}
{\rm Tr}\, \left(  M^2_\eta  \right) &=& 
{\rm Tr}\, \left(  {M^2_\eta}  \right)_{\rm exp}
\nonumber \\ 
{\rm det}\, \left( M^2_\eta \right) &=& 
{\rm det}\, \left( {M^2_\eta} \right)_{\rm exp}
\label{trace_det_eq}
\end{eqnarray}
 We identify the lighest two $\eta$'s 
predicted by our model (i.e. $\eta_1$ and 
$\eta_2$) with $\eta 
(547)$ and $\eta'(958)$ with 
experimental masses \cite{RPP}:
\begin{eqnarray}
m^{\rm exp.}[\eta (547)] &=& 547.853 \pm 
0.024\, {\rm 
MeV},\nonumber \\
m^{\rm exp.}[\eta' (958)] &=& 957.66 \pm 0.24 
\, {\rm 
MeV}.
\end{eqnarray}
 However, for the two heavier $\eta$'s 
that our model predicts (i.e. $\eta_3$ and
$\eta_4$) there 
are
several experimental candidates below 2 GeV 
with masses \cite{RPP}:
\begin{eqnarray}
m^{\rm exp.}[\eta (1295)] &=& 1294 \pm 4\, {\rm 
MeV},\nonumber \\ 
m^{\rm exp.}[\eta (1405)] &=& 1409.8 \pm 2.5 \, 
{\rm 
MeV},
\nonumber \\
m^{\rm exp.}[\eta (1475)] &=& 1476 \pm 4\, {\rm 
MeV},\nonumber \\ 
m^{\rm exp.}[\eta (1760)] &=& 1756 \pm 9 \, 
{\rm 
MeV}.
\end{eqnarray}
We consider all six possible choices for 
identifying $\eta_3$ and
$\eta_4$ with two of the above four 
experimental candidates.   This leads to six 
scenarios given in table \ref{etascenarios}.
Equations (\ref{trace_det_eq}) result in a 
quadratic equation for $\gamma_1$ for which the 
discriminant versus $m_\pi(1300)$ is plotted 
for all six scenarios in Fig. \ref{Fig_dscm}.
We see that for $A_3/A_1$ = 20 the scenarios 
1, 2 and 4 are completely ruled out, whereas 
for  $A_3/A_1$ = 30 up to $m_\pi(1300) \approx$ 
1.25 GeV all six scenarios are possible.   

\begin{figure}
\begin{center}
\vskip 1cm
%-----------------------------------------
\epsfxsize = 6cm
 \epsfbox{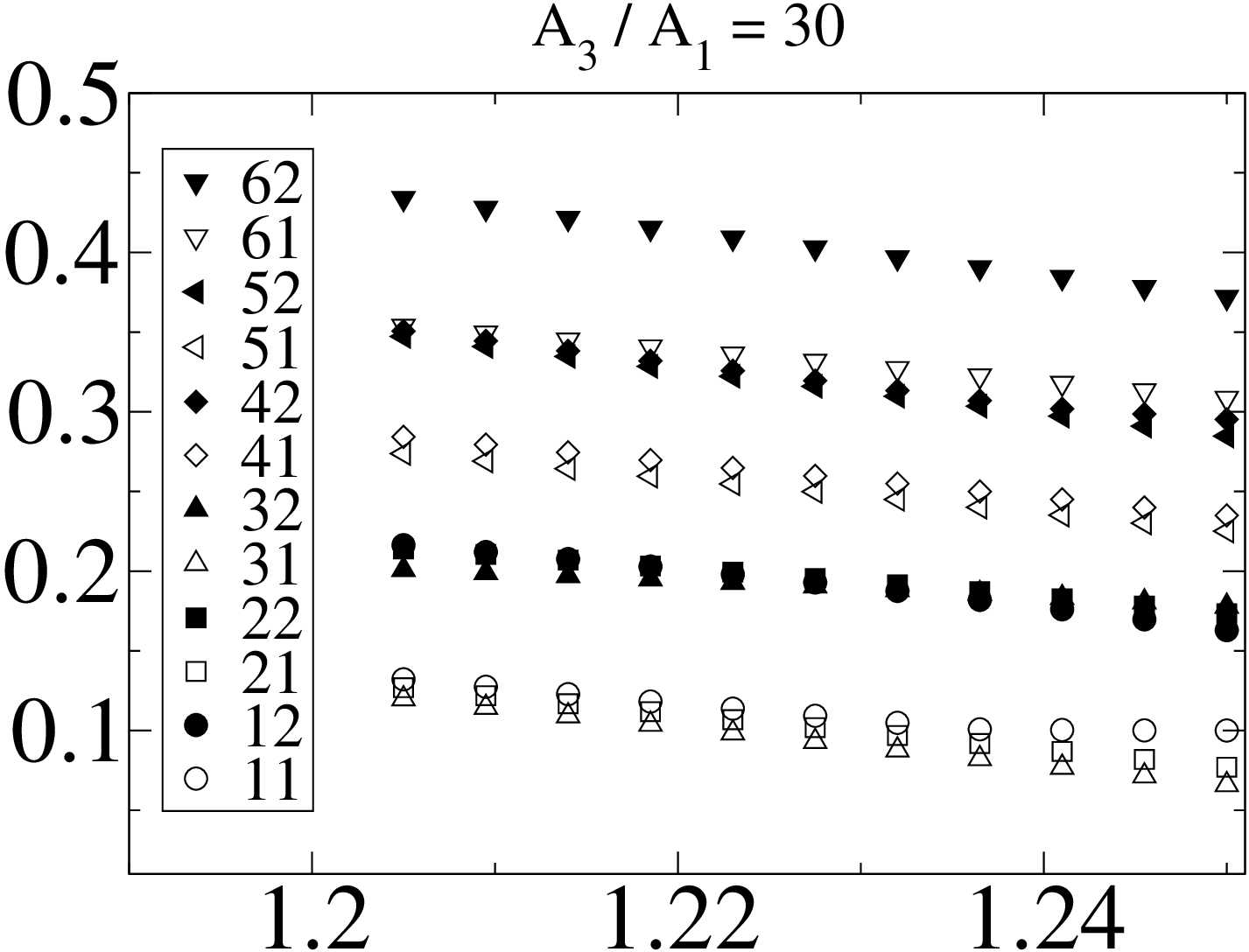}
%-----------------------------------------
\end{center}
\caption[]{%
$\chi_{sl}$
vs	   
$m[\pi(1300)]$ (GeV).
}
\label{Fig_chi_vs_mpip}
\end{figure}

Therefore, for a given scenario there are two 
solutions for $\gamma_1$ and $c_3$, and 
consequently two sets of predictions for the 
four $\eta$ masses.    
We measure the  
goodness of each solution by the smallness of 
the following quantity:
\begin{equation}
\chi_{sl} = 
\sum_{k=1}^4
{
 {\left| m^{\rm theo.}_{sl}(\eta_k)  -  
  m^{\rm exp.}_{s}(\eta_k)\right|}    
                 \over 
  m^{\rm exp.}_{s}(\eta_k)  
}
\end{equation}  
in which $s$ corresponds to the scenario 
(i.e. $s= 1 \cdots 6$) and 
$l$ corresponds to the solution number 
(i.e. $l= 1,2$).   The quantity $\chi_{sl} 
\times 100$ gives the overall percent 
discrepancy between our theoretical prediction 
and experiment.   For the six scenarios and 
the two solutions for each scenario, 
$\chi_{sl}$ is plotted versus $m_\pi(1300)$ in 
Fig. \ref{Fig_chi_vs_mpip}.    Clearly, 
scenario 3 is favored over the range of 
$m_\pi(1300)$.   

\begin{figure}
\begin{center}
\vskip 1cm
%-----------------------------------------
\epsfxsize = 6cm
 \epsfbox{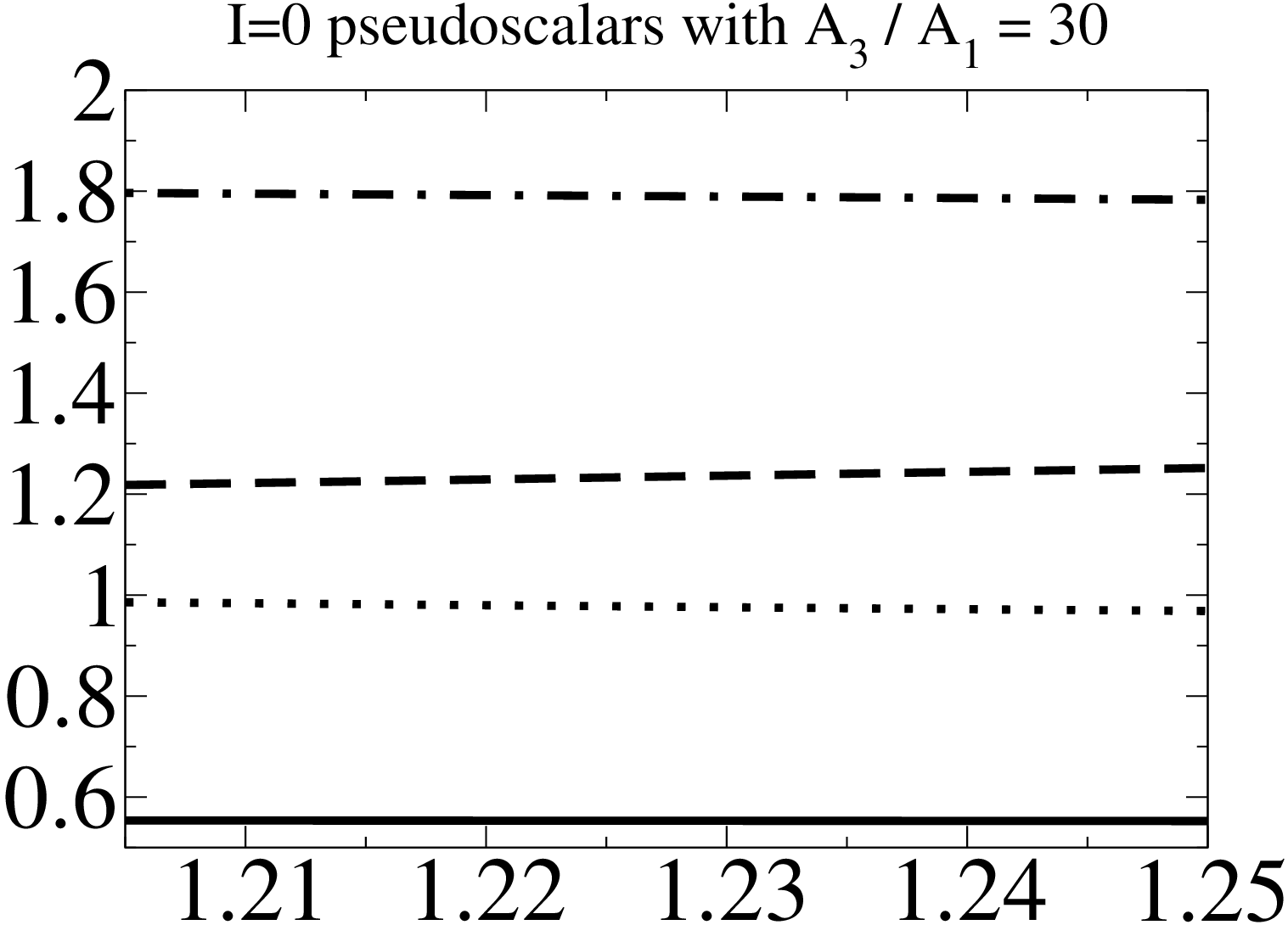}
%-----------------------------------------
\end{center}
\caption[]{%
Predicted $\eta$ masses
vs	   
$m[\pi(1300)]$ (GeV).
}
\label{Fig_etamasses_vs_mpip}
\end{figure}

  Here, we
present our predictions for the best
fitting scenario 3 with
$A_3/A_1$ = 30.
(Of course, the other scenarios with
different values of $A_3/A_1$ 
may also be of some interest for a more
detailed look at additional properties of the
system of the four etas).
  In the present case, the $\eta$
masses are shown in Fig.  
\ref{Fig_etamasses_vs_mpip} and we can clearly
see a reasonable range of masses:  The first
mass is around 550 MeV (consistent with
identifying it with $\eta(547)$), the second
mass is in the range of 970 - 986 MeV
(consistent with identifying it with
$\eta(985)$), the third mass is in the range of
1218 - 1250 MeV (consistent with identifying it
with $\eta(1295)$) and the fourth mass is
around 1780 - 1790 MeV (consistent with
identifying it with $\eta(1760)$).

\begin{figure}
\begin{center}
\vskip 1cm
%-----------------------------------------
\epsfxsize = 6cm
\epsfbox{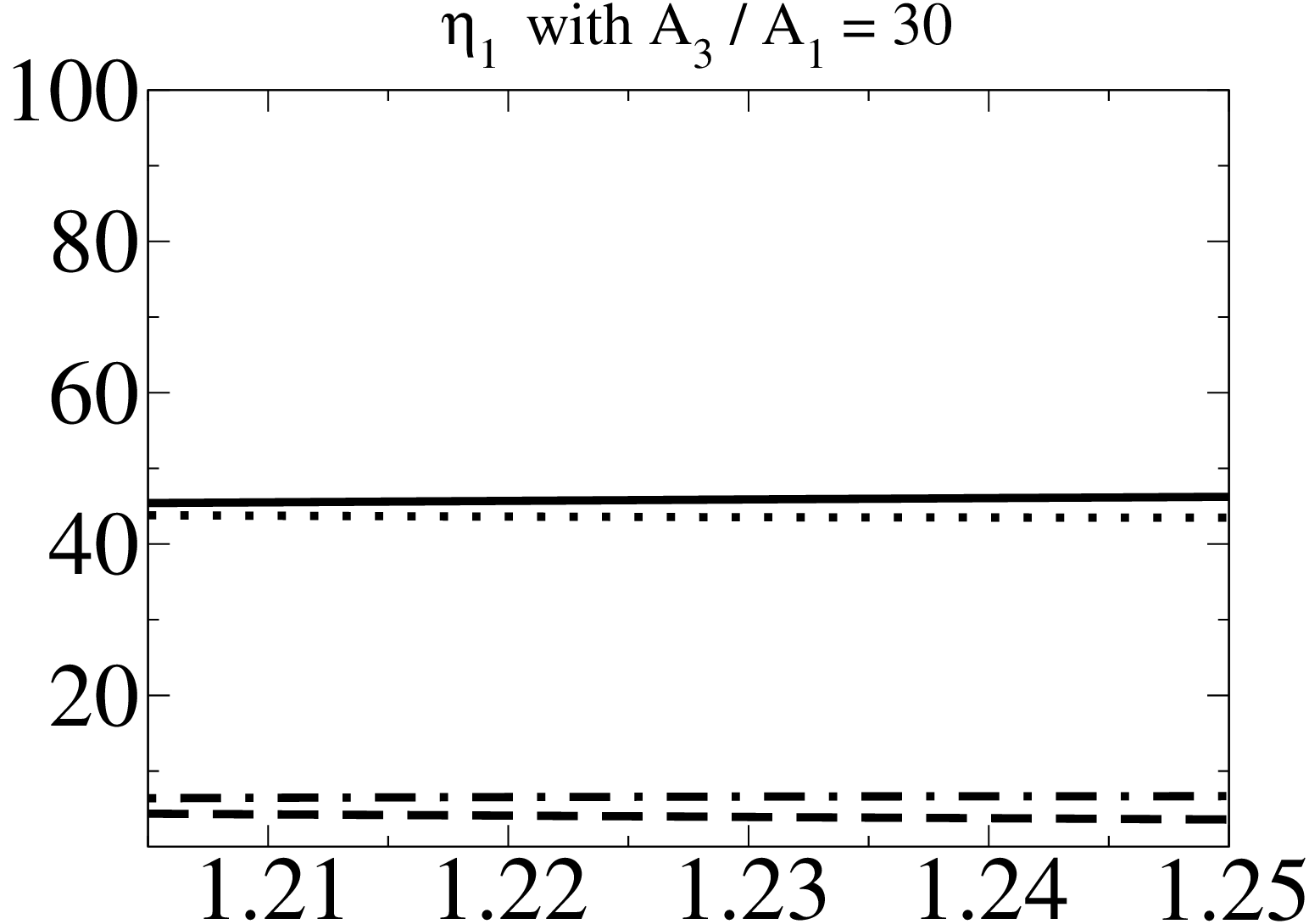}
\hskip 1cm
\epsfxsize = 6cm
 \epsfbox{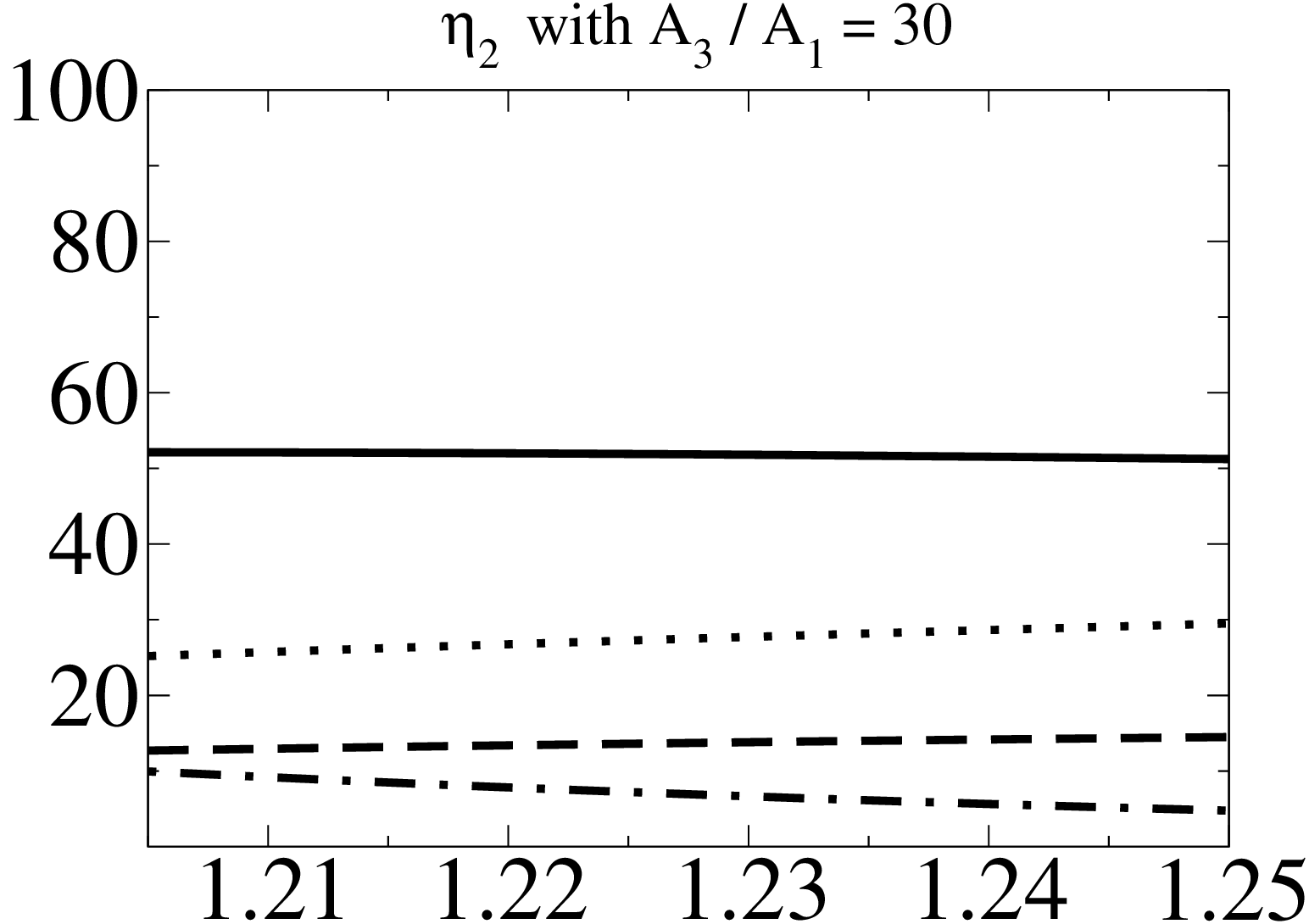}
\vskip 1cm
%-----------------------------------------
\epsfxsize = 6cm
\epsfbox{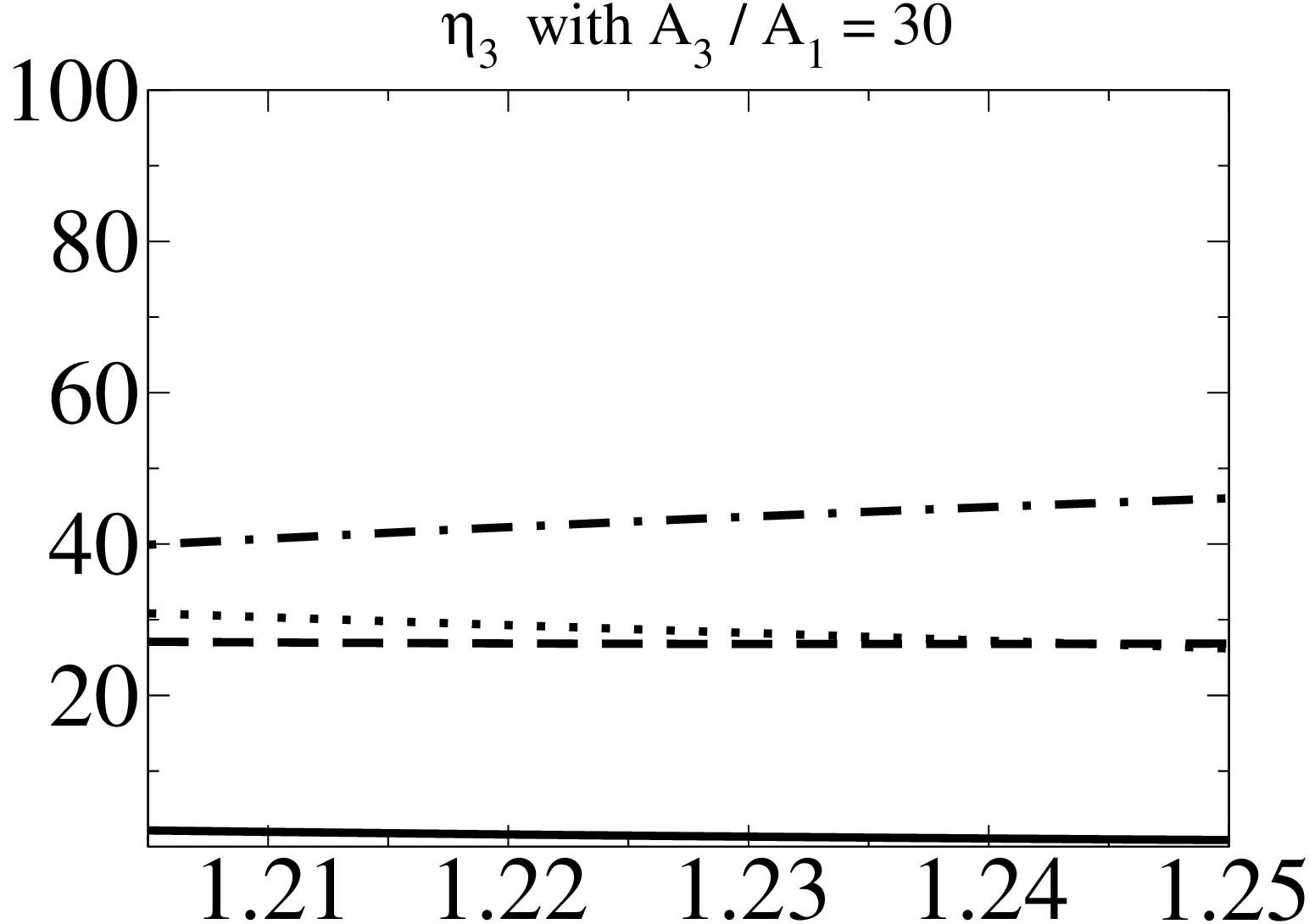}
\hskip 1cm
\epsfxsize = 6cm
 \epsfbox{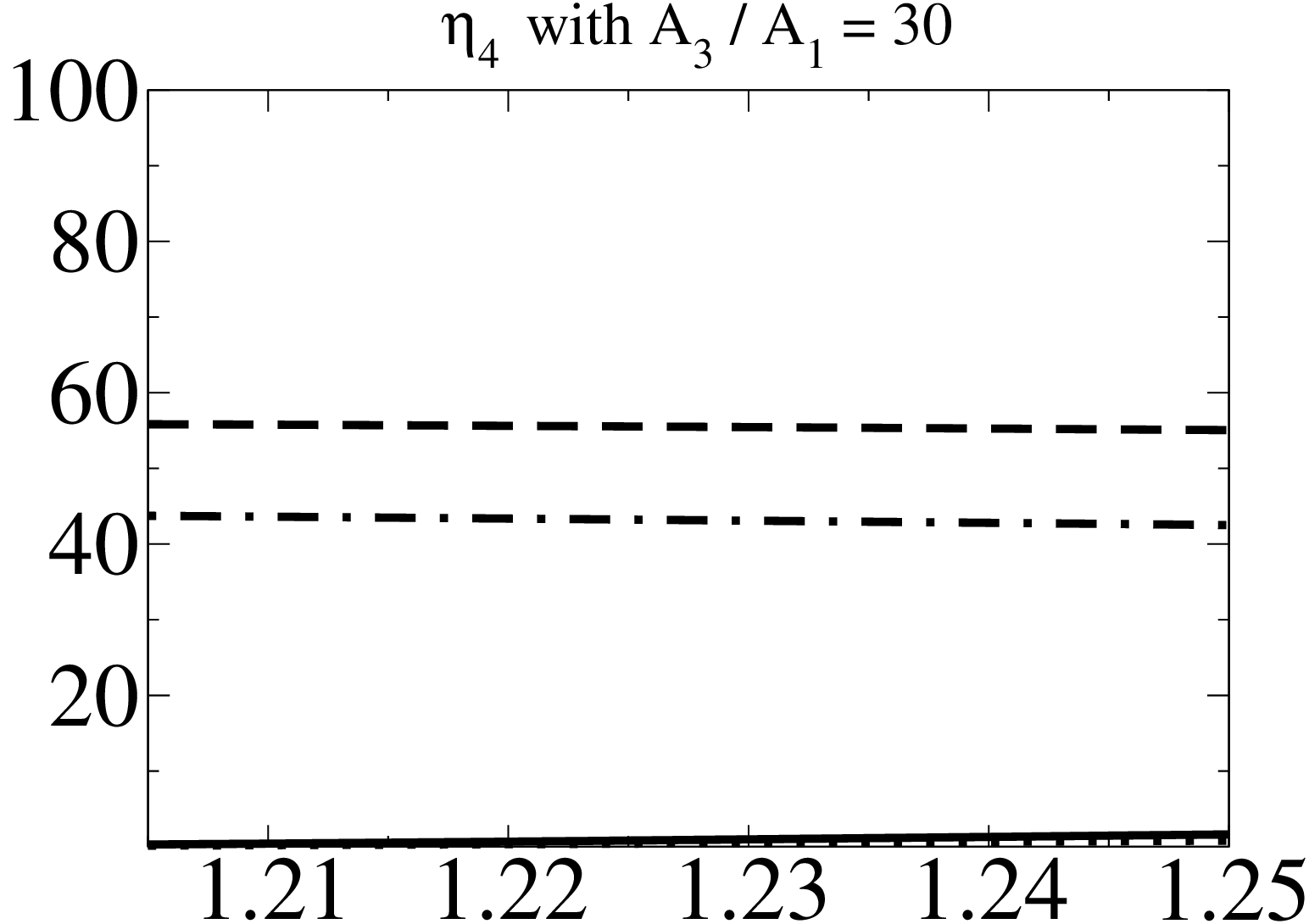}
%-----------------------------------------
\end{center}
\caption[]{%
Predicted quark contents of isoscalar 
pseudoscalar  
mesons vs	   
$m[\pi(1300)]$ (GeV).   The components 
[described in Eq. (\ref{etafourbasis})] are:
$\eta_a$ (solid line), 
$\eta_b$ (dotted line), 
$\eta_c$ (dashed line) and  
$\eta_d$ (dotted-dashed line). 
}
\label{Fig_etacomps}
\end{figure}

The quark contents of the $\eta$'s are given in
Fig. \ref{Fig_etacomps} and show a clear
difference compared to those of the I=0 
scalars.  The
first two states ($\eta(547)$ and $\eta(985)$)  
are dominantly of two quark nature whereas the
two heavier states ($\eta(1295)$ and
$\eta(1760)$) are mainly four-quark states.

\begin{table}[htbp]
\begin{center}
\begin{tabular}{c||c}
\hline \hline
Scenario &
$\eta$'s
\\
\hline
\hline
1 &
$\eta(547), \eta(958), \eta(1295), \eta(1405)$ 
\\ \hline
2 &
$\eta(547), \eta(958), \eta(1295), \eta(1475)$ 
\\
\hline
3 & $\eta(547), \eta(958), \eta(1295), 
\eta(1760)$ 
\\
\hline
4 & $\eta(547), \eta(958), \eta(1405),  
\eta(1475)$ 
\\
\hline
5 & $\eta(547), \eta(958), \eta(1405), 
\eta(1760)$ 
\\
\hline
6 & $\eta(547), \eta(958), \eta(1475), 
\eta(1760)$ 
\\
\hline
\hline
\end{tabular}
\end{center}
\caption[]{The six scenarios for identifying 
the predictions of our model for four $\eta$s.}
\label{etascenarios}
\end{table}

\section{Conclusions and Discussion}

    The results obtained \cite{results} seem to
provide support for the picture of light
scalars having a predominant content of two
quarks and two antiquarks while the heavier
scalars appear to be made mainly from one quark
and one antiquark, as one would expect from p
wave states in the non relativistic quark
model. The predicted scalar and 
pseudoscalar masses were seen to be
 reasonably consistent 
with the experimental candidates listed in 
\cite{RPP}.

    While it appears a little unusual to think of
say, the ordinary pion, as having some 
 four quark content when treated in the
 effective Lagrangian framework, that is in fact 
the standard picture in the parton model approach
to QCD. In the case of the two scalar nonets, the
mass ordering itself naturally suggested such a picture.
This picture was then inherited by the pseudoscalars 
when we chose to describe the scalars via a linear
sigma model. There does not seem to be any problem with 
our treatment of the lighter pseudoscalar nonet. One
might initially worry that the well
 established ``current algebra"
theorem for low energy pion pion scattering
 could get altered in the case of such
 a more complicated
pion. However, we showed in
\cite{2FJS07} and  \cite{3FJS07} that there
is only a small effect for this theorem. 
The situation concerning the heavier (mostly four
quark) pseudoscalar nonet which appears in our model is
neither so clear experimentally nor theoretically.
 Evidently this seems 
a fruitful direction for further investigation.
The two new $\eta$s which appear in this model are
welcome from an experimental standpoint and the
mixing with an expected glueball in this energy 
range is another interesting future topic.
There is of course a similar glueball expected
 in the scalar isoscalar channel.

    In addition to the ``global" analysis
 of the mass
spectrum and substructure presented here,
the ``local" topic of the detailed structure
of the pion and kaon scattering is also of great
interest. Recently progress has been made in this area,
strongly supporting the existence of both the sigma
 and kappa mesons,
by using (\cite{CCL} and \cite{DM})
 the chiral perturbation theory approach
combined with some dispersion theory
 results \cite{R}. Even though the present model
is rather complicated, the fact that it is chiral
symmetric means that we are guaranteed to get
essentially the starting results of the
 chiral perturbation theory approach near
the appropriate thresholds. In fact, since the
light scalar fields are included, the 
starting results are even closer to the modern
ones just mentioned. Thus we plan to next investigate
the meson meson scattering scatterings in 
a number of different channels and at energies
 away from threshold. We are encouraged
to go further by the results \cite{BFMNS01}
obtained some time
ago by unitarizing the scattering amplitudes
 computed
using a single chiral SU(3) nonet model.

    Actually, we have already started on this 
work and would like to briefly mention a result 
for the lowest lying scalar meson since it 
gives an idea of the accuracy to be expected
for the masses we have presented in this paper.
A more detailed analysis of the scattering
 amplitudes
will be presented elsewhere.

% we have investigated in detail 
%the
%prediction of this model for the real part of
%$\pi\pi$ scattering amplitude.  We have found
%that, expectedly, the inclusion of SU(3)  
%symmetry breaking significantly improves the
%prediction of (K-matrix unitarized) scattering
%amplitude up to about 1 GeV.  This clearly
%provides further support for the consistency of
%our systematic treatment of the 
%generalized linear sigma model Lagrangian.

The mass spectrum of the light scalars  
receives considerable unitarity corrections due 
to 
low-energy rescattering effects.
The masses obtained above appear as tree level
quantities in the effective Lagrangian under discussion.
Especially in the case of the scalars the physical
states are rather broad and will appear as poles
 in the scattering of two pseudoscalar mesons.
A simple way to estimate the scattering
amplitude is to first
compute the tree level scalar partial wave scattering
amplitude and then unitarize it
 by using the K-matrix method.
This is equivalent to an earlier approach \cite{AS94}
 and amounts to replacing the tree level
 amplitude, $T_{tree}$ by,
\begin{equation}
T=\frac{T_{tree}}{1-iT_{tree}}.
\label{kmatrix}
\end{equation}
Following this procedure the sigma pole at 742 MeV
discussed above appears
 in the unitarized
pion scattering amplitude at
$z\equiv M^2
-iM\Gamma$ with
\begin{equation}
M=477\, {\rm  MeV}, \,\, \Gamma= 504\,  {\rm 
MeV}.
\end{equation}
This is of the same order as in \cite{CCL}.
 Such scattering
calculations should be performed to find the
``actual" mass and width parameters for
all the scalars in the present model.

\section*{Acknowledgments} \vskip -.5cm
We are happy to thank A. Abdel-Rehim, D. Black, M. Harada,
S. Moussa, S. Nasri, N.W. Park, A.D. Polosa, F. Sannino
and M.N. Shahid for many helpful
related discussions.
The work of A.H.F. has been partially supported by the
NSF Award 0652853.
The work of J.S. is supported in part by the U. S. DOE under
Contract no. DE-FG-02-85ER 40231.

\appendix
\section{Parameter Evaluation}
  
    By using the following formulas consecutively,
 it is 
possible to determine all the parameters
 of the model,
one at a time, from the experimental inputs.
 The procedure is 
a more complicated version of the zero quark mass
 case given in Appendix 
B of \cite{2FJS07} and the degenerate non-zero
quark mass case given in the 
Appendix of \cite{3FJS07}.

We first use the I=1 scalar and pseudoscalar squared
mass matrices to get the parameters,  
 
\begin{eqnarray}
2 d_2 &=& {  {m_a^2 m_{a'}^2  - m_\pi^2\, m_{\pi'}^2} \over
{m_a^2 +m_{a'}^2 - m_\pi^2 - m_{\pi'}^2} }
\nonumber \\
(\alpha_3\, e_3^a)^2 &=&\frac{1}{64}( (m_a^2 -
m_{a'}^2)^2  - [4 d_2 - (m_a^2
+m_{a'}^2)]^2)
\label{d2ande3}
\end{eqnarray}

The minimum equation then yields the ratio of the four
quark to the two quark ``condensate":

\begin{equation}
{\beta_1\over \alpha_1} = -{{2 (\alpha_3
e_3^a)} \over d_2}.
\label{beoveral}
\end{equation}

Next, the $\pi-\pi'$ mixing angle is found
from the diagonalization of $(M_\pi^2)$: 

\begin{equation} 
{\rm cos}\, 2\theta_\pi =
{
   {4\, d_2 - m_\pi^2 - m_{\pi'}^2}
  \over
   {
    \sqrt{
           64\, (\alpha_3\, e_3^a)^2
           + 16\, d_2^2
           - 8\, d_2 \, (m_\pi^2 + m_{\pi'}^2)
           + (m_\pi^2 + m_{\pi'}^2)^2
         }
   }
}.
\end{equation}

Then we get, from the formula for the pion
decay constant \cite{FJS05},

\begin{equation} 
 \alpha_1 =
{1\over 2}\,
{
 {F_\pi}
   \over
 { {\rm cos}\, \theta_\pi -
\left({\beta_1\over \alpha_1}\right) \,
    {\rm sin}\, \theta_\pi
 }
},  
\end{equation}

which then also gives $\beta_1$ using 
Eq.(\ref{beoveral}) above.

To go beyond this point we introduce the 
strange to non-strange quark mass ratio, $A_3/A_1$
as an input parameter. We may then solve for
the quantity $e_3^a\, \alpha_1$ by using the two
minimum equations:
\begin{eqnarray}
A_1 &=&
2\, \beta_3 \, (e_3^a\, \alpha_1)  + 2\, \beta_1 \, 
(e_3^a \, \alpha_3) -  c_2 \, \alpha_1 +
2 \, c_4^a \, \alpha_1^3
\nonumber \\  
A_3 &=&
4 \, \beta_1 \, ( e_3^a \, \alpha_1) -
c_2 \, \alpha_3 + 2 \, c_4^a \, \alpha_3^3
\label{A1A2}
\end{eqnarray}

Now, the rest of the parameters may be obtained
by the consecutive use of the following equations:

\begin{eqnarray}
c_4^a &=& {1 \over {8\, \alpha_1^2}}
\left[
m_a^2 + m_{a'}^2
- m_\pi^2 - m_{\pi'}^2
- 16 \,{ {(\alpha_1 \, e_3^a)^2} \over d_2}
\right]   
\nonumber \\
4 \, c_2 &=&
16 \, \alpha_1^2 \, c_4^a +
4 \, d_2 -
m_\pi^2 -  m_{\pi'}^2
- m_a^2 -  m_{a'}^2
\nonumber \\
e_3^a &=& { {(e_3^a \, \alpha_1)} \over
           \alpha_1}
\nonumber \\
\alpha_3 &=& { {(e_3^a \, \alpha_3)} \over
           e_3^a}
\nonumber \\
\beta_3 &=&
{
  {\alpha_1\, \beta_1}
          \over
    \alpha_3  
}
\nonumber \\
(M^2_K)_{11} &=&
m_\pi^2 + m_{\pi'}^2 - 2\, d_2
+
{1\over 2}\,
(m_a^2 + m_{a'}^2 - m_\pi^2 - m_{\pi'}^2)\,
\left(  
{
  {(e_3^a \, \alpha_3)^2}\over
  {(e_3^a \, \alpha_1)^2 }
} 
  -
{
  {(e_3^a \, \alpha_3) }
  \over {(e_3^a \, \alpha_1)}
}
\right)
\nonumber \\
&&
-
{8\over d_2}\,
\left[
(\alpha_3\, e_3^a)^2
-
(\alpha_1\, e_3^a)^2
\right]
\nonumber \\
(M^2_K)_{12} &=&
- 4\, (e_3^a \, \alpha_1)
\nonumber \\
(M^2_K)_{22} &=&
2 \, d_2
\nonumber \\
{\rm cos}\, 2\theta_K &=&
{
   {(M^2_K)_{22} - (M^2_K)_{11}}
  \over
   {
    \sqrt{
           4\, (M^2_K)_{12}^2 +
          \left[(M^2_K)_{22} - (M^2_K)_{11}
          \right]^2
          }
   }
}
\nonumber \\
{F_K \over F_\pi} &=&
{ 
 {
  \left(
         1 + {{(e_3^a \, \alpha_3)}
                     \over
              {(e_3^a \, \alpha_1)}
             }
  \right) \cos \theta_K -
         {\beta_1 \over \alpha_1}
         \left[ 1 +
            {{(e_3^a \, \alpha_1)}
                     \over
              {(e_3^a \, \alpha_3)}
             }
         \right] \sin \theta_K
 }
\over
 {
    2 \,  \cos \theta_\pi -
    2 \, \left( {\beta_1 \over
                 \alpha_1}
          \right) \sin \theta_\pi
 }
}
\label{lagpara}
\end{eqnarray}

With $m[\pi(1300)]$ = 1.215 GeV and $A_3/A_1$ = 30, these
parameters are given in table \ref{T_10param}. In this case,
the rotation matrices that related the flavor bases to the
physical states are given in (\ref{2by2rot}) and 
(\ref{4by4rots}).

\begin{table}[htbp]
\begin{center}
\begin{tabular}{c|c}
\hline \hline
$c_2 ({\rm GeV}^2)$    & 1.62 $\times 10^{-1}$ 
\\
$d_2  ({\rm GeV}^2)$   & 6.30 $\times 
10^{-1}$\\
$e_3^a  ({\rm GeV})$   & $-1.68$\\
$c_4^a $               & 47.0  \\
$\alpha_1  ({\rm GeV})$  & 6.06 $\times 
10^{-2}$\\
$\alpha_3  ({\rm GeV})$  & 7.68 $\times 
10^{-2}$\\
$\beta_1  ({\rm GeV})$   & 2.49 $\times 
10^{-2}$\\
$\beta_3  ({\rm GeV})$   & 1.96 $\times 
10^{-2}$\\
$A_1 ({\rm GeV}^3)$   & 6.66 $\times 10^{-4}$
\\
$A_3 ({\rm GeV}^3)$   & 2.00 $\times 10^{-2}$
\\
\hline
\end{tabular}
\end{center}
\caption[]{
Calculated Lagrangian parameters:$c_2$, $d_2$, 
$e_3^a$, $c_4^a$
and vacuum values: $\alpha_1$, 
$\alpha_3$, $\beta_1$ and $\beta_3$, with
$m[\pi(1300)]$ = 1.215 GeV and $A_3/A_1$ =
30.
}
\label{T_10param}
\end{table}

\begin{equation}
(R_\pi^{-1})  =
\left[
\begin{array}{cc}
0.923 & $0.385$ \\
-$0.385$ & 0.923\\
\end{array}
\right], \hspace{.3cm}
(R_K^{-1})  =
\left[
\begin{array}{cc}
0.925 & $0.379$ \\
-$0.379$ & 0.925\\
\end{array}
\right], \hspace{.3cm}
(L_a^{-1})  =
\left[
\begin{array}{cc}
$0.493$ & -$0.870$ \\
$0.870$    & 0.493\\
\end{array}
\right], \hspace{.3cm}
(L_\kappa^{-1})  =
\left[
\begin{array}{cc}
0.284 & -$0.959$ \\
$0.959$    & 0.284\\
\end{array}
\right].
\label{2by2rot}
\end{equation}

\begin{equation}
(L_o^{-1})  =
\left[
\begin{array}{cccc}
0.601  &  0.199 &  0.600  &    0.489 \\ 
-0.107   &  0.189 &  0.643  &     -0.735  \\
0.790   &  -0.050&  -0.391  &   -0.470  \\ 
0.062 &  -0.960 &   0.272  &   -0.019\\
\end{array}
\right]
\label{4by4rots}
\end{equation}

\section{$M^2_\eta$}

The elements of the symmetric matrix $M^2_\eta$
are given by: 

\begin{eqnarray}
\left( M^2_\eta \right)_{11} &=&
\left(
16 \, c_4 \, \alpha_1^6
{{\beta_1}}^{2} + 16\,  c_4 \,
{{\alpha_1}}^{5} {\alpha_3} \,
{\beta_1} \,
{\beta_3} + 4\, {c_4} \,
{{\alpha_1}}^{4}{{\alpha_3}}^{2}
{{\beta_3}}^{2}-16\,{e_3}\,
{{\alpha_1}}^{4}{{\beta_1}}^{2}
{\beta_3} - 16\,{e_3}\,
{{\alpha_1}}^{3}{\alpha_3}\,{\beta_1}\,
{{\beta_3}}^{2}
\right.
\nonumber
\\
&&
- 4\,{e_3}\,
{{\alpha_1}}^{2}{{\alpha_3}}^{2}
{{\beta_3}}^{3}
- 8\,{c_2}\,
{{\alpha_1}}^{4}{{\beta_1}}^{2}-8\,
{c_2}\,{{\alpha_1}}^{3}
{\alpha_3}\,{\beta_1}\,{\beta_3}-2\,
{c_2}\,{{\alpha_1}}^{2}
{{\alpha_3}}^{2}{{\beta_3}}^{2} - 16\,
{c_3}\,{{\gamma_1}}^{2}
{{\alpha_1}}^{2}{{\beta_1}}^{2}
\nonumber
\\
&&
-32\,
{c_3}\,{{\gamma_1}}^{2}{
\alpha_1}\,{\alpha_3}\,{\beta_1}\,{\beta_3}
-16\,{c_3}\,{{
\gamma_1}}^{2}{{\alpha_3}}^{2}{{\beta_3}}^{2}
-32\,{c_3}\,{
\gamma_1}\,{{\alpha_1}}^{2}{{\beta_1}}^{2}-32\,
{c_3}\,{\gamma_1}\,{\alpha_1}\,
{\alpha_3}\,{\beta_1}\,{\beta_3}
\nonumber
\\
&&
\left.
-16\,{c_3}\,
{{\alpha_1}}^{2}{{\beta_1}}^{2}
\right) /
\left(
\left( 2\, {\alpha_1}\,{\beta_1}+
{\alpha_3} \, {\beta_3} \right)
^{2}{{\alpha_1}}^{2}
\right)
 \end{eqnarray}

\begin{eqnarray}
\left( M^2_\eta \right)_{12} &=&
\left(
-4\,\sqrt {2} \left( 4\,{e_3}\,{{
\alpha_1}}^{3}{\alpha_3}\,{{
 \beta_1}}^{3} + 4\,{e_3}\,{{
\alpha_1}}^{2}{{\alpha_3}}^{2}{{
\beta_1}}^{2}{\beta_3}+{e3}\,{\alpha_1}\,{{
\alpha_3}}^{3}{\beta_1}\,{{\beta_3}}^{2}+4\,
{c3}\,{{\gamma_1}}^{2}{{\alpha_1}}^{2}{{\beta_1}}
^{2}+
\right.
\right.
\nonumber \\
&&
\left.
\left.
4\,{c_3}\,{{\gamma_1}}^{2}{\alpha_1}\,{\alpha_3}\,
{\beta_1}\,{\beta_3}+4\,{c3}\,{\gamma_1}\,
{{\alpha_1}}^{2}{{\beta_1}}^{2}+2\,{c_3}\,{\gamma_1}
\,{\alpha_1}\,{\alpha_3}\,{\beta_1}\,{\beta_3}+
2\,{c_3}\,{\gamma_1}\,{{\alpha_3}}^{2}{{\beta_3}}
^{2}+2\,{c_3}\,{\alpha_1}\,{\alpha_3}\,{\beta_1}\,
{\beta_3}
\right)
\right)
\nonumber \\
&&
/
 \left(
 \left(
2\,{\alpha_1}\,{\beta_1}+{\alpha_3}\,{\beta_3}
 \right) ^{2}{\alpha_1}\,{\alpha_3}
 \right)
\end{eqnarray}
\begin{eqnarray}
\left( M^2_\eta
 \right)_{13} &=&
\left(
-16\,
{e_3}\,{{\alpha_1}}^{2}{\alpha_3}\,{{\beta_1}}^
{2}
-
16\,
{e_3}\,{\alpha_1}\,{{\alpha_3}}^{2}{\beta_1}\,
{\beta_3}-4\,{e_3}\,{{\alpha_3}}^{3}{{\beta_3}}^{2}
-16\,{c_3}\,{{\gamma_1}}^{2}{\alpha_1}\,{\beta_1}
-16\,{c_3}\,{{\gamma_1}}^{2}
{\alpha_3}\,{\beta_3}
\right.
\nonumber \\
&&
\left.
+ 16\,{c_3}\,{\gamma_1}\,{\alpha_3}\,{\beta_3}
+ 16\,{c_3}\,{\alpha_1}\,{\beta_1}
\right)
/
\left(
 \left(
2\,{\alpha_1}\,{\beta_1}+{\alpha_3}\,{\beta_3}
 \right) ^{2}
\right)
\end{eqnarray}

\begin{eqnarray}
\left( M^2_\eta \right)_{14} &=&
\left(
-4\,\sqrt {2} \left(
4\,{e_3}\,{{\alpha_1}}^{4}{{\beta_1}}^{2}+
4\,{e_3}\,{{\alpha_1}}^{3}{\alpha_3}\,{\beta_1}\,
{\beta_3}+
{e_3}\,{{\alpha_1}}^{2}{{\alpha_3}}^{2}{{\beta_3}}^
{2}+2\,{c_3}\,{{\gamma_1}}^{2}{\alpha_1}\,{\alpha_3}
\,{\beta_1}
\right.
\right.
\nonumber \\
&&
\left.
\left.
+2\,
{c_3}\,{{\gamma_1}}^{2}{{\alpha_3}}^{2}{\beta_3}
-2\,{c_3}
\,{\gamma_1}\,{{\alpha_3}}^{2}{\beta_3}-2\,
{c_3}\,{\alpha_1
}\,{\alpha_3}\,{\beta_1} \right)
\right)
\nonumber \\
&&
/
\left(
2\,{\alpha_1}\,{\beta_1}+{\alpha_3}\,{\beta_3}
 \right) ^{2}{\alpha_1}
\end{eqnarray}

\begin{eqnarray}
\left( M^2_\eta \right)_{22} &=&
\left(
16\,{c_4}\,{{\alpha_1}}^{2}{{\alpha_3}}^{4}
{{\beta_1}}^{2}+16\,{c_4}\,{\alpha_1}\,
{{\alpha_3}}^{5}{\beta_1}\,{\beta_3}+4
\,{c_4}\,{{\alpha_3}}^{6}{{\beta_3}}^{2}
-8\,{c_2}\,{{\alpha_1}}^{2}{{\alpha_3}}^{2}
{{\beta_1}}^{2}-8\,{c_2}\,{\alpha_1}\,
{{\alpha_3}}^{3}{\beta_1}\,{\beta_3}
\right.
\nonumber\\
&&
\left.
-2\,{c_2}\,{{
\alpha_3}}^{4}{{\beta_3}}^{2}-32\,{c_3}\,
{{\gamma_1}}^{2}{{\alpha_1}}^{2}{{\beta_1}}^{2}
-32\,{c_3}\,{\gamma_1}\,{\alpha_1}\,{\alpha_3}
\,{\beta_1}\,{\beta_3}-8\,{c_3}\,
{{\alpha_3}}^{2}{{\beta_3}}^{2}
\right)
\nonumber \\
&&
/
\left(
\left( 2\,{\alpha_1}\,{\beta_1}+
{\alpha_3}\,{\beta_3}
 \right) ^{2}{{\alpha_3}}^{2}
\right)
\end{eqnarray}

\begin{eqnarray}
\left( M^2_\eta \right)_{23} &=&
\left(
-4\,\sqrt {2} \left( 4\,{e_3}\,
{{\alpha_1}}^{2}{\alpha_3}\,{{
\beta_1}}^{2}+4\,{e_3}\,{\alpha_1}\,
{{\alpha_3}}^{2}{\beta_1}\,{\beta_3}+{e_3}\,
{{\alpha_3}}^{3}{{\beta_3}}^{2}+4\,{
c_3}\,{{\gamma_1}}^{2}{\alpha_1}\,{\beta_1}-4\,
{c_3}\,{\gamma_1}\,{\alpha_1}\,{\beta_1}
\right.
\right.
\nonumber \\
&&
\left.
\left.
+2\,{c_3}\,
{\gamma_1}\,{\alpha_3}\,{\beta_3}-2\,{c_3}\,{\alpha_3}
\,{\beta_3}
 \right) {\alpha_1}
\right)
/
\left(
 \left(
2\,{\alpha_1}\,{\beta_1}+{\alpha_3}\,{\beta_3}
 \right) ^{2}{\alpha_3}
\right)
\end{eqnarray}

\begin{eqnarray}
\left( M^2_\eta \right)_{24} &=&
\left(
-8\, \left( {\gamma_1}-1 \right)  \left(
2\,{\gamma_1}\,{\alpha_1}\,{\beta_1}+{\alpha_3}
\,{\beta_3}
\right) {c_3}
\right)
/
 \left(
2\,{\alpha_1}\,{\beta_1}+{\alpha_3}\,{\beta_3}
 \right) ^{2}
\end{eqnarray}

\begin{eqnarray}
\left( M^2_\eta \right)_{33} &=&
\left(
8\,{d_2}\,{{\alpha_1}}^{2}{{\beta_1}}^{2}+8\,{d_2}
\,{\alpha_1}\,{\alpha_3}\,{\beta_1}\,{\beta_3}+
2\,{d_2}\,{{\alpha_3}}^{2}{{\beta_3}}^{2}-16\,{c_3}
\,{{\gamma_1}}^{2}{{\alpha_1}}^{2}+32\,{c_3}\,
{\gamma_1}\,{{\alpha_1}}^{2}-16\,{c_3}\,
{{\alpha_1}}^{2}
\right)
\nonumber \\
&&
/
 \left(
2\,{\alpha_1}\,{\beta_1}+{\alpha_3}\,{\beta_3}
 \right) ^{2}
\end{eqnarray}

\begin{eqnarray}
\left( M^2_\eta \right)_{34} &=&
\left(
-8\, \left( {\gamma_1}-1 \right) ^{2}\sqrt
{2}{c_3}\,{\alpha_1}
\,{\alpha_3}
\right)
/
\left(
2\,{\alpha_1}\,{\beta_1}+{\alpha_3}\,{\beta_3}
 \right) ^{2}
\end{eqnarray}

\begin{eqnarray}
\left( M^2_\eta \right)_{44} &=&
\left(
8
\,{d_2}\,{{\alpha_1}}^{2}{{\beta_1}}^{2}
+8\,{d_2}\,{\alpha_1}\,{\alpha_3}\,{\beta_1}\,
{\beta_3}+2\,{d_2}\,{{\alpha_3}}^{2}{{\beta_3}}
^{2}-8\,{c_3}\,{{\gamma_1}}^{2}{{\alpha_3}}^{2}
+16\,{c_3}\,{\gamma_1}\,{{\alpha_3}}^{2}-8\,{c_3
}\,{{\alpha_3}}^{2}
\right)
\nonumber \\
&&
/
 \left(
2\,{\alpha_1}\,{\beta_1}+{\alpha_3}\,{\beta_3}
 \right) ^{2}
\end{eqnarray}

The basis states for the above matrix 
are:
\begin{eqnarray}
\eta_a&=&\frac{\phi^1_1+\phi^2_2}{\sqrt{2}},
\nonumber  \\
\eta_b&=&\phi^3_3,
\nonumber    \\
\eta_c&=&  \frac{\phi'^1_1+\phi'^2_2}{\sqrt{2}},
\nonumber   \\
\eta_d&=& \phi'^3_3.
\label{etafourbasis}
\end{eqnarray}

With $m[\pi(1300)]$ = 1.215 GeV and $A_3/A_1$ = 
30, these paprameters are given in table 
\ref{T_2param}.

\begin{table}[htbp]
\begin{center}
\begin{tabular}{c|c}
\hline \hline
$c_3 ({\rm GeV}^4)$ & $-3.78 \times 
10^{-4}$ \\
$\gamma_1$          & 5.27 $\times 10^{-3}$ 
\\
\hline
\end{tabular}
\end{center}
\caption[]{
Calculated parameters: $c_3$ and $\gamma_1$.
}
\label{T_2param}
\end{table}

In this case, the rotation matrix for I=0
pseudoscalars becomes:

\begin{equation}
(R_o^{-1})  =
\left[
\begin{array}{cccc}
-0.675  &  0.661 &  -0.205  &    0.255 \\ 
0.722   &  0.512 &  -0.363  &     0.291  \\
-0.134   &  -0.546&  -0.519  &   0.644  \\ 
0.073 &  0.051 &   0.746  &   0.660\\
\end{array}
\right]
\label{mms}
\end{equation}

\end{document}